\documentclass[%
 reprint,
 amsmath,amssymb,
 prl,
aps,
]{revtex4-1}

\usepackage{graphicx}
\usepackage{dcolumn}
\usepackage{bm}
\usepackage{float}
\usepackage{physics}
\usepackage{bm,amsmath}
\usepackage{natbib}
\usepackage{xr}
\newcommand{\units}[1]{\ensuremath{\mathrm{#1}}}

\newcommand{\amount}[2]{\ensuremath{#1\:\units{#2}}}

\newcommand{\beginsupplement}{%
        \setcounter{table}{0}
        \renewcommand{\thetable}{S\arabic{table}}%
        \setcounter{figure}{0}
        \renewcommand{\thefigure}{S\arabic{figure}}%
        \setcounter{equation}{0}
        \renewcommand{\theequation}{S\arabic{equation}}
        \setcounter{section}{0}
        \renewcommand{\thesection}{S\Roman{section}}
        }
        
\usepackage{hyperref}

\begin{document}


\title{Coherent control and spectroscopy of a semiconductor quantum dot Wigner molecule}

\author{J. Corrigan$^{1}$} 
\email{jbaer4@wisc.edu}
\author{J. P. Dodson$^{1}$}%
\author{H. Ekmel Ercan$^{1}$}
\author{J. C. Abadillo-Uriel$^{1}$}
\author{Brandur Thorgrimsson$^{1}$}
\author{T. J. Knapp$^{1}$}
\author{Nathan Holman$^{1}$}
\author{Thomas McJunkin$^{1}$}
\author{Samuel F. Neyens$^{1}$}
\author{E. R. MacQuarrie$^{1}$}
\author{Ryan H. Foote$^{1}$}
\author{L. F. Edge$^{2}$}
\author{Mark Friesen$^{1}$}
\author{S. N. Coppersmith$^{1,3}$}
\author{M. A. Eriksson$^{1}$}

\affiliation{$^{1}$University of Wisconsin, Madison, WI 53706}
\affiliation{$^{2}$HRL Laboratories, LLC, 3011 Malibu Canyon Road, Malibu, CA 90265}
\affiliation{$^{3}$University of New South Wales, Sydney, Australia}

\date{\today}

\begin{abstract}

Multi-electron semiconductor quantum dots have found wide application in qubits, where they enable readout and enhance polarizability. However, coherent control in such dots has typically been restricted to only the lowest two levels, and such control in the strongly interacting regime has not been realized. Here we report quantum control of eight different resonances in a silicon-based quantum dot. We use qubit readout to perform spectroscopy, revealing a dense set of energy levels with characteristic spacing far smaller than the single-particle energy. By comparing with full configuration interaction calculations, we argue that the dense set of levels arises from Wigner-molecule physics.
\end{abstract}

\pacs{Valid PACS appear here}
\maketitle

Semiconductor quantum dots containing more than one electron have extremely desirable properties for constructing and operating qubits. For single spin qubits, manipulating electrons above closed shells makes electric-field driving more effective \cite{Yang:2020p350, Leon:2020p797}, and certain qubits like quantum dot hybrid qubits \cite{Kim:2014p70} rely on multielectron filling to define the qubit states themselves. Two-electron eigenstate energies are particularly important for quantum dot qubits, since singlet-triplet splittings result in Pauli spin-blockade \cite{Ono:2002p1313}, which provides a readout mechanism for singlet-triplet qubits \cite{Petta:2005p2180,Shulman:2012p202,Reed:2016p110402}, exchange based qubits \cite{Gaudreau:2011p54, Medford:2013p050501}, and quantum dot hybrid qubits \cite{Shi:2012p140503, Kim:2014p70}.  Pauli blockade can also be used to read out single spin qubits \cite{Li:2018p7}, which is useful for high temperature operation \cite{Vandersypen:2017p34,Petit:2020p355, Yang:2020p350}. When the characteristic interaction energy between electrons becomes larger than the orbital confinement energy, electronic states develop correlations and localize, leading to the formation of Wigner molecules \cite{Bryant:1987p1140, Yannouleas:1999p5325, Reimann:2000p8108, Reusch:2001p113313,Rontani:2006p124102,Ghosal:2007p085341}.  Imaging of localization in Wigner molecules has been achieved using scanning electronic \cite{Shapir:2019p870} and near-field optical \cite{Mintairov:2018p195443} methods.  The lowest-lying excited states in such molecules have been studied using both optical \cite{Kalliakos:2008p467, Singha:2010p246802} and transport spectroscopy \cite{Ellenberger:2006p126806, Kristinsdottir:2011p041101}, and the latter method has been used to observe a reduction in symmetric-antisymmetric orbital splittings \cite{Pecker:2013p576}.  Thus, while the transition to Wigner-type localization is known to reduce the gap between the ground and first orbital excited state, the impact on higher lying states has not been observed in experiment, and quantum control of such states has not been explored.

In this Letter we report pulsed microwave coherent control and spectroscopy of an electrostatically-confined semiconductor double quantum dot in the Wigner-molecule regime.  We report coherent Rabi control of eight distinct resonances ranging in frequency from 3.3 to 8.3~GHz, corresponding to energies far smaller than the single-particle confinement energy.  With Ramsey spectroscopy, we map the energy as a function of double-dot detuning for two of these resonances. Using full configuration interaction (FCI) calculations, we argue that the origin of this dense manifold of electronic states is strong correlations and Wigner molecule physics.  Time-domain simulations of the Rabi experiments are used to explain the evolution of the Rabi oscillations as a function of detuning energy.  The full set of experimental spectroscopy results can be fit by a simple model consisting of a set of two-electron states in the right quantum dot tunnel coupled to the lowest lying state in the left dot.

\begin{figure}[H]
\centering
  \includegraphics[trim = 10 426 170 20, clip, width = 17.8cm]{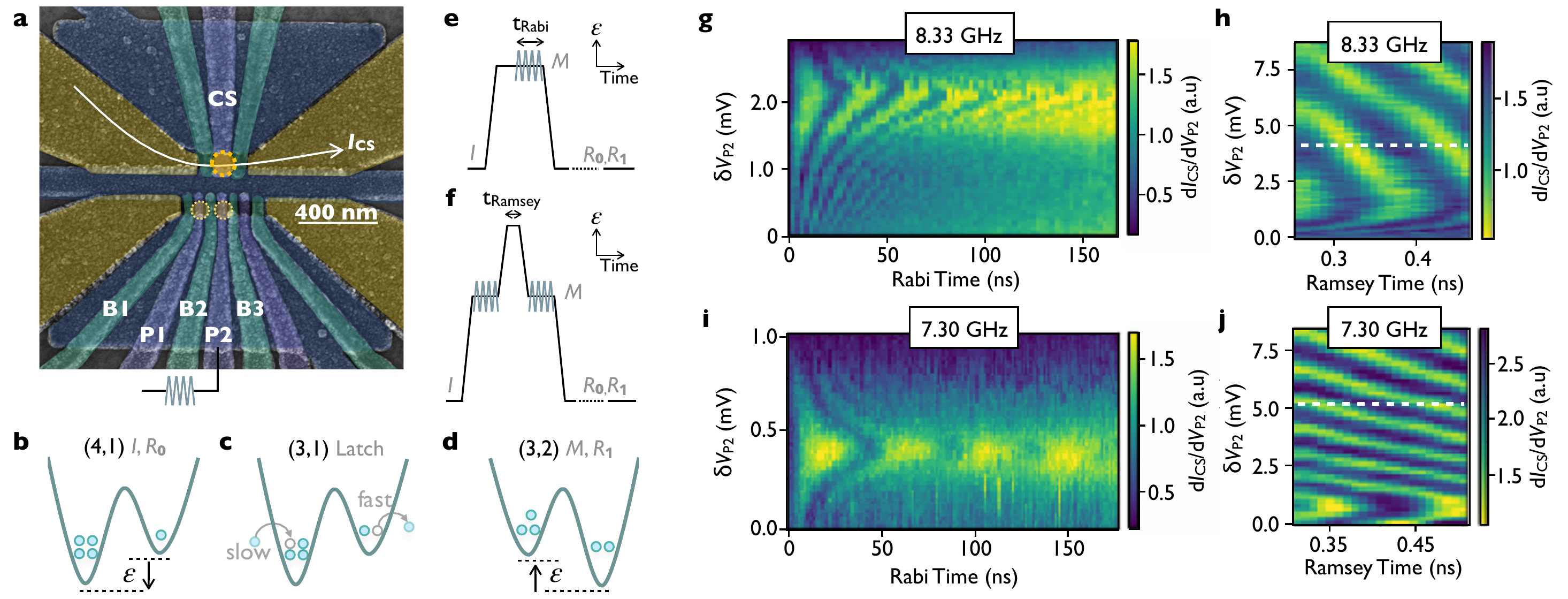}
  \caption{ 
(a) A false-color micrograph of a device lithographically identical to that measured here. Quantum dots are formed under P1 and P2. A current ${I_{\mathrm CS}}$ flows under the dot controlled by gate CS and is used to detect the electron occupation of the P1 and P2 dots. (b) The ground state for the 5-electron system at negative detuning $\varepsilon$ is the (4,1) charge configuration.  It is used for initialization $I$ and during readout $R_{0}$. (c) The tunnel rates to the left and right reservoirs are tuned such that (3,1) state is metastable, allowing for latched readout. (d) The ground state for the 5-electron system at positive detuning such that the double dot is in the (3,2) charge configuration, used for manipulation $M$ and readout $R_{1}$. (e) Rabi pulse sequence used in this work. (f) Ramsey pulse sequence used in this work, comprised of two $\pi /2$ pulses on either side of a detuning pulse.
(g-j) Rabi and detuned-Ramsey measurements of two coexisting states. (g,h) Rabi and Ramsey oscillations using $f_{\mathrm{R}}$ = \amount{8.33}{GHz}. (i,j) Rabi and Ramsey oscillations using $f_{\mathrm{R}}$ = \amount{7.30}{GHz}, taken at the same device tuning but different detuning value from (g,h). For the Rabi measurements in (g, i), $\delta V_{\mathrm{P2}}$ corresponds to a shift of the entire pulse sequence, while for the Ramsey measurements in (h,j),  $\delta V_{\mathrm{P2}}$ corresponds to the height of the detuned dc pulse between the microwave bursts. Dashed white lines in (h,j) denote the value of $\delta V_{\mathrm{P2}}$ for which the Ramsey detuning is zero.}

  \label{fgr:SEM}
\end{figure}

Fig.~\ref{fgr:SEM}(a) describes the quantum dots used here, which form in an undoped Si/SiGe heterostructure with three layers of overlapping gates \cite{Zajac:2015p223507}.  Fabrication details can be found in Ref.~\cite{Dodson:2020preprint}.  While the device is capable of hosting three dots, here we make use of the two dots under gates P1 and P2, accumulating the rightmost dot as part of the electron reservoir. We operate the double quantum dot (DQD) with a total of five electrons near the (4,1)-(3,2) anticrossing, as shown schematically in Fig.~1(b-d). Tunnel rates between the two dots and to the left and right reservoirs are set by gates B1, B2, and B3.  Charge sensing is performed with dot CS, and its current is measured using a two-stage cryogenic HEMT amplifier \cite{Tracy:2016p063101} mounted on a  separate printed circuit board (PCB) connected to the sample PCB by stainless steel coax.

We initialize at setting $I$, shown in Fig.~\ref{fgr:SEM}(b), into the (4,1) ground state, which has a large splitting between the ground and first excited states. We ramp the detuning across the interdot transition line to a manipulation point ($M$) in positive detuning where we perform microwave pulse sequences. Rabi and detuned Ramsey pulses at $M$ drive coherent rotations between states in the (3,2) DQD. To perform readout, we adiabatically ramp back across the interdot transition line: while the (3,2) ground state maps to the (4,1) ground state ($R_{0}$), the (3,2) excited states maintain their charge configuration ($R_{1}$). Latched measurement  \cite{Studenikin:2012p233101} is used to enhance readout: an electron in the right-hand dot of the (3,2) 
\onecolumngrid

\twocolumngrid

\noindent charge state at $R_{1}$ rapidly tunnels into the right reservoir to form the metastable (3,1) charge state before slowly returning to the (4,1) ground state. The latch duration is determined by the left barrier, which is tuned to have a long tunnel time. 

Figure~\ref{fgr:SEM}(e-j) demonstrate coherent spectroscopy of two different transitions of the DQD, using the methods from Ref.~\cite{Thorgrimsson:2017p32}. Fig.~\ref{fgr:SEM}(e,f) shows the Rabi and Ramsey control pulses; the Rabi pulse consists of one continuous microwave drive of frequency $f_{\mathrm{R}}$, while the Ramsey pulse has two $\pi/2$ microwave pulses of frequency $f_{\mathrm{R}}$ surrounding a dc detuning ramp. The resulting Rabi and Ramsey oscillations are shown in Fig.~\ref{fgr:SEM}(g,h) and Fig.~\ref{fgr:SEM}(i,j) for $f_{\mathrm{R}}$ of  \amount{8.33}{GHz} and \amount{7.30}{GHz}, respectively. The vertical axis $\delta V_{\mathrm{P2}}$ determines the detuning, and the centers of the Rabi chevrons in Fig.~\ref{fgr:SEM}(g,i) correspond to the detuning values where $f_{\mathrm{R}}$ is resonant with the transition energy.
Ramsey fringes like those in Fig.~\ref{fgr:SEM}(h,j) allow measurement of the energy difference as a function of detuning (Fig.~\ref{fgr:datashift} in Ref.~\cite{Suppl}).

\begin{figure}[H]
\centering
\includegraphics[trim = 10 320 511 30, clip, width = 8.5cm]{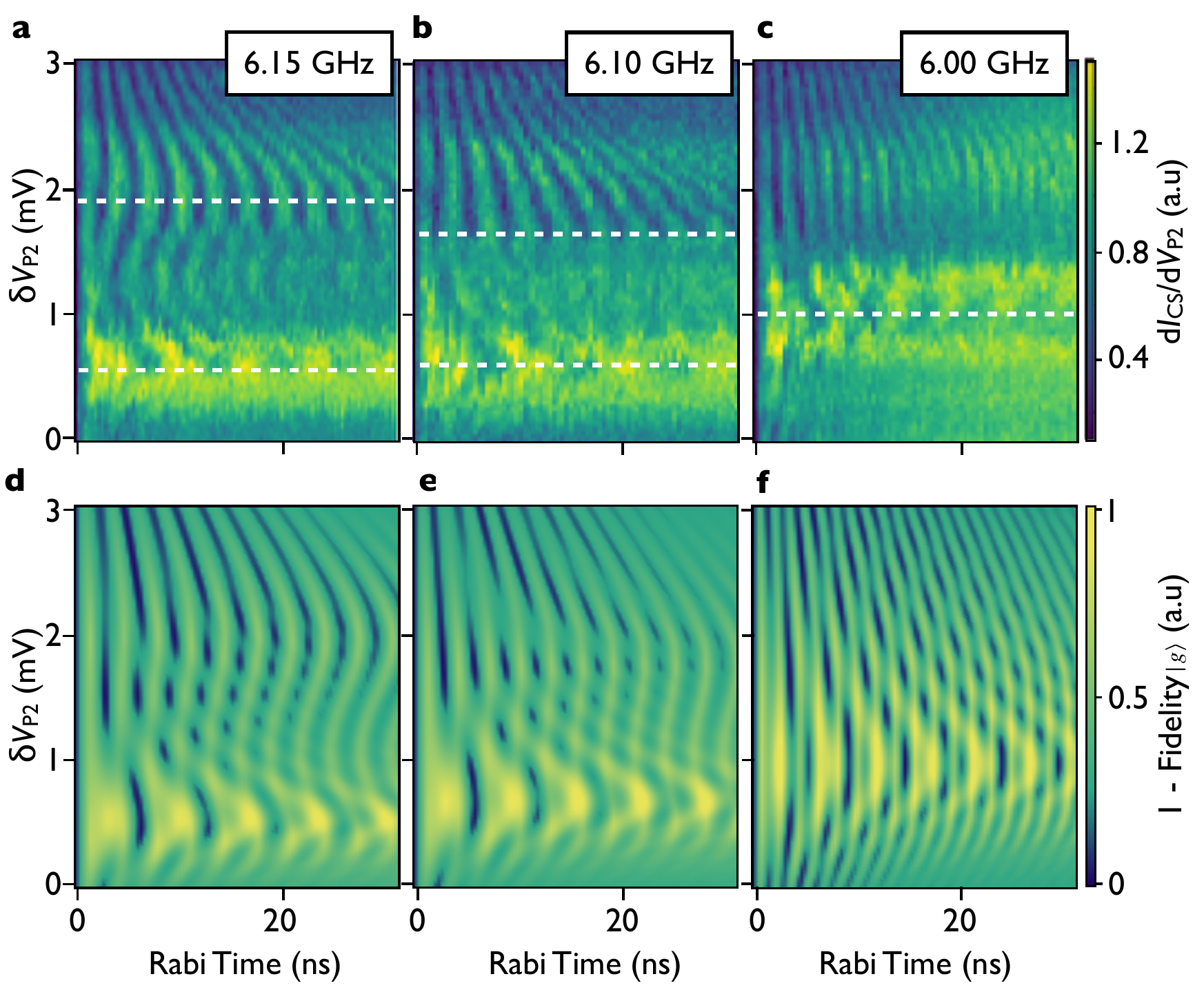}
  \caption{
  (a) Rabi oscillations with $f_{\mathrm{R}}$ = \amount{6.15}{GHz}, with the centers of two on-resonance oscillations marked by a dashed line. (b) Rabi oscillations with  $f_{\mathrm{R}}$ = \amount{6.10}{GHz}.  Note that the two resonances move closer together along the detuning axes, as compared with (a). (c) Rabi oscillations with  $f_{\mathrm{R}}$ = \amount{6.00}{GHz}, where the on-resonance locations have completely merged. (d-f) Simulated Rabi oscillations corresponding to (a-c) using the simplified model described in Sec.~S2 of \cite{Suppl}.
}

  \label{fgr:manyres}
\end{figure}

Figure~\ref{fgr:manyres} shows Rabi oscillations with two distinct resonances visible in the same plot. 
In Fig.~\ref{fgr:manyres}(a), the Rabi drive frequency $f_{\mathrm{R}}$ = \amount{6.15}{GHz}, and the detunings corresponding to on-resonance oscillations are indicated by the white dashed lines in the figure. As the driving frequency is reduced to \amount{6.10}{GHz} (Fig.~\ref{fgr:manyres}(b)), the centers of the Rabi patterns move closer together, and in Fig.~\ref{fgr:manyres}(c) the oscillations 
overlap at $\delta V_{\mathrm{P2}}$ = \amount{1}{mV}. Numerical simulations of these oscillations are shown in Fig.~\ref{fgr:manyres}(d-f), 
%
\noindent using the model described in Sec.~S2 of Ref.~\cite{Suppl}. A key feature of these two oscillations is the difference in width as a function of $\delta V_{\mathrm{P2}}$; this behavior is reproduced in the theoretical model by different slopes for the respective energy dispersions, where a flatter slope corresponds to wider oscillations. The unusual merging of the resonances is reproduced in the model with a level-crossing, which we discuss further below.
  
Figures~\ref{fgr:SEM} and \ref{fgr:manyres} above report four resonances as a function of the gate voltages defining the quantum dot. Additional Rabi measurements over a wide range of frequencies are included in Fig.~\ref{fgr:additionaldoubledata},~\ref{fgr:allrabipoints} of Ref.~\cite{Suppl}. In total this data demonstrates Rabi driving of eight distinct transitions below \amount{10}{GHz}, an unusual density of resonances that cannot be described by non-interacting two-electron physics. We must therefore consider how electron-electron interactions influence the excited energy level spectrum.

\begin{figure}[H]
\centering
\includegraphics[trim = 35 390 200 30, clip, width = 17.8cm]{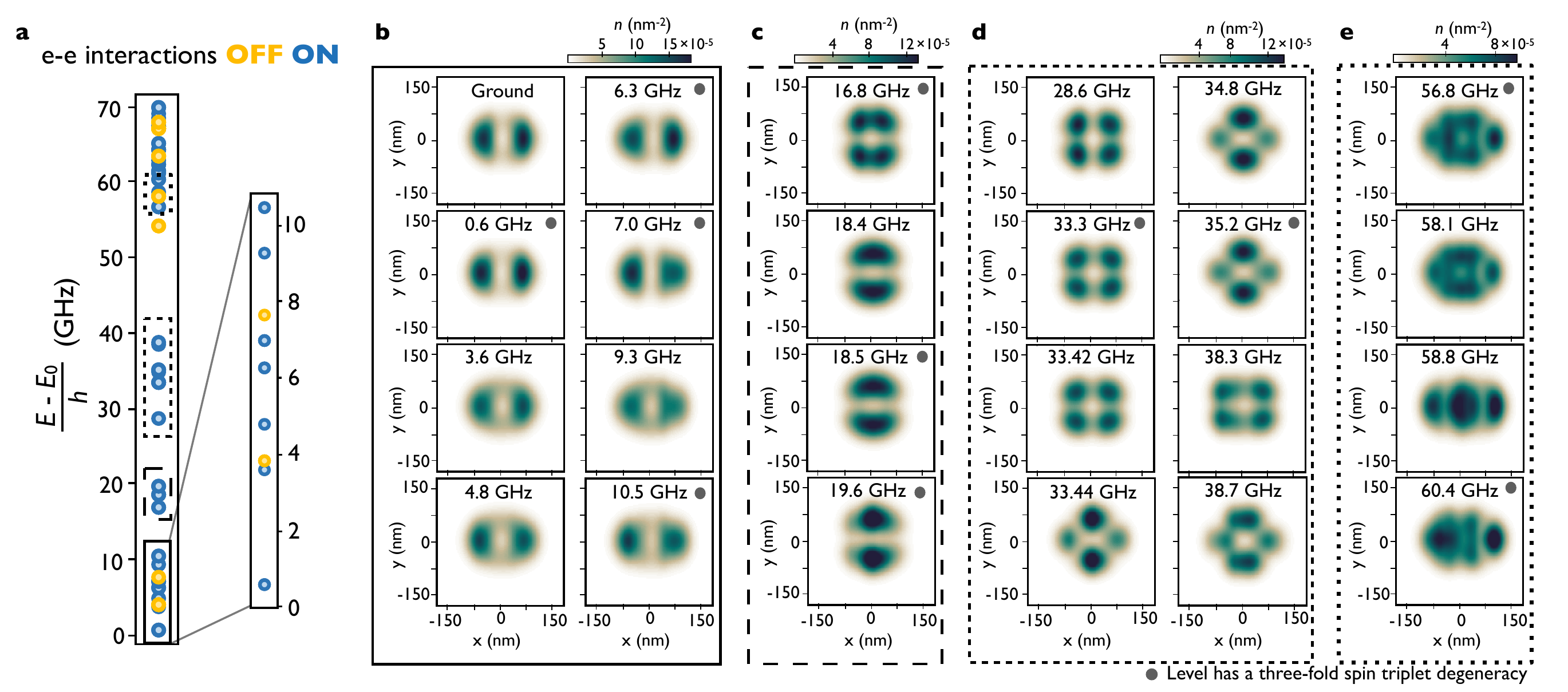}
  \caption{
Interaction effects. (a) Excited state energies for a two-electron parabolic dot described by the energy $\hbar \omega_{x}$, with $\omega_{x}/2\pi$ = \amount{59.2}{GHz} ($\hbar \omega_{y}$ = 1.07 $\hbar \omega_{x}$), calculated using FCI with (blue) and without (yellow) interactions.  (b-e) Electron density for the blue levels in (a) with each panel corresponding to a boxed cluster of levels with the corresponding border style. }

  \label{fgr:theory}
\end{figure}

Here, we use FCI methods to compute the energy levels in the two-electron quantum dot \cite{Rontani:2006p124102}. We find that the best correspondence between theory and experiment is achieved by assuming a valley splitting of 3.81~GHz and an orbital confinement $E_{\mathrm{orb}}/h$ = \amount{59.2}{GHz}. It is interesting to first consider the noninteracting case, for which the lowest singlet state is constructed from the two lowest single-electron orbitals, while the lowest triplet is constructed from the ground and first excited states. We obtain these and the other low-lying, non-interacting two-electron eigenstates by simply turning off the Coulomb interactions in the FCI code, obtaining the yellow points in Fig.~\ref{fgr:theory}(a). In contrast, the fully interacting case is shown in blue
(Sec.~S4 in Ref.~\cite{Suppl} and Ref.~\cite{ErcanPrep}). For the fully interacting case, when interactions are larger than valley or orbital excitation energies, the eigenstates are composed of contributions from many single-electron states. In Fig.~\ref{fgr:theory}(a), we observe the emergence of level manifolds; for eigenstates within a given manifold, single-particle contributions and electron correlation effects can be very similar in magnitude, and opposite in sign, resulting in a much smaller energy difference than in the non-interacting limit (Sec.~S5
in Ref.~\cite{Suppl}).  For the parameters considered here, strong interactions yield a dense set of levels, with a total of 19 excited states below \amount{50}{GHz}, instead of the two states obtained in the noninteracting limit. Interactions also suppress the energy splitting between the two lowest energy states to below 1 GHz. 

Wigner molecules with localized electrons are known to arise in systems with a high ratio  $R_{\mathrm{W}}$  = $E_{\mathrm{ee}}$ / $E_{\mathrm{orb}}$ between the electron-electron interaction energy $E_{\mathrm{ee}}$ and the energy of the lowest quantum dot orbital excitation energy $E_{\mathrm{orb}}$. Typical estimates for the formation of Wigner molecules are of order $R_{\mathrm{W}}$ = 1.5 
\cite{Ellenberger:2006p126806, Kalliakos:2008p467, Singha:2010p246802, Pecker:2013p576, Mintairov:2018p195443}. Here, using $E_{\mathrm{orb}}$, we find $R_{\mathrm{W}}$ = 12.74, consistent with the formation of a Wigner molecule (Sec.~S5 in Ref.~\cite{Suppl}). The estimate for $E_{\mathrm{orb}}$ used to compute $R_{\mathrm{W}}$ also yields a dot radius of \amount{40}{nm}, derived from the classical turning point $\sqrt{\hbar/(m_{t} \omega_{x})}$, giving an 80~nm diameter that is consistent with a quantum dot situated in a \amount{90}{nm} channel underneath a gate \amount{70}{nm} wide.  We note that, if we use any of the two-electron excited state energies reported here to estimate the dot diameter in a noninteracting electron model, it would give a dot diameter over 200~nm, much larger than the size expected from the physical dimensions of the electrostatic gates in the 
\onecolumngrid

\twocolumngrid
\clearpage
\newpage
\mbox{~}
\clearpage
\newpage

 \noindent device
 
 Figure~\ref{fgr:theory}(b-e) shows electron density plots corresponding to the blue energy levels in Fig.~\ref{fgr:theory}(a), separated into four different manifolds, as indicated by the solid, dashed, and dotted bounding boxes. It is clear that these Wigner molecule eigenstates cannot be easily constructed from low-energy, single-electron orbital states. This manifold structure also helps to explain the energy spectrum in Fig.~\ref{fgr:theory}(a), since the gaps in the calculated spectrum correspond to changes in the spatial pattern of electron localization. Energy separations within a given manifold can be attributed to slight differences in orbital configurations and to varying contributions from valley excitations. Larger energy gaps between the manifolds correspond to characteristic changes in the electron localization.

\begin{figure}[H]
\centering
  \includegraphics[trim = 20 160 560 40, clip, width = 8.5cm]{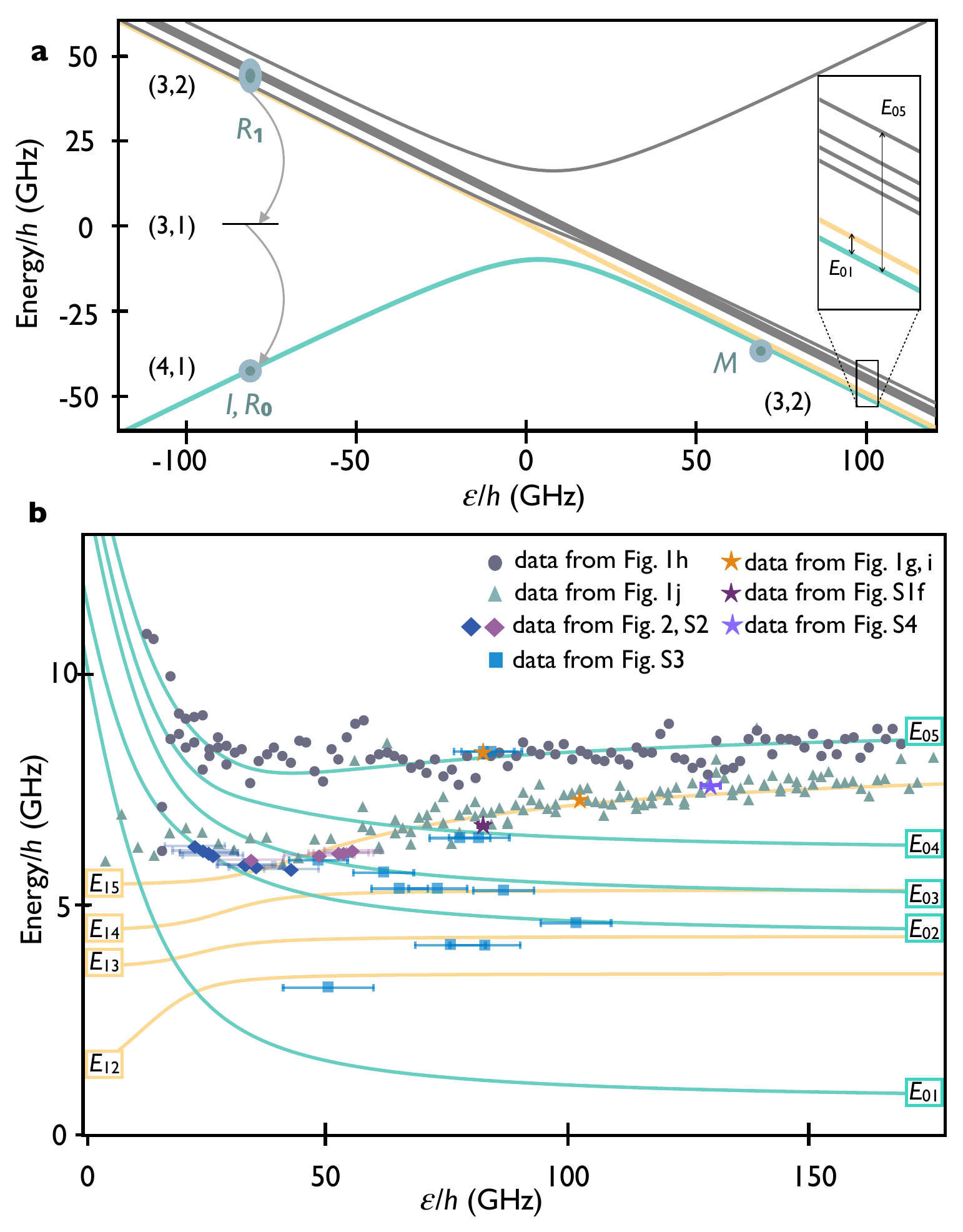}
  \caption{
(a) Energy eigenvalues of a model Hamiltonian motivated by the electron interaction effects reported here. Measurement locations from Fig.~1 are indicated. (b) Excited state spectra of the model Hamiltonian. The teal lines are the energy differences between the ground state and energy level n, defined in (a); the yellow lines are differences between the first excited level and the same energy levels in (a). Symbols plotted correspond to frequencies extracted from experiment, as described in the legend. }

  \label{fgr:alldata}
\end{figure}

Figure~\ref{fgr:alldata}(a) shows the eigenvalues of a simple Hamiltonian (Eq.~\ref{eq:Hamiltonian} of Ref.~\cite{Suppl}) motivated by the energy levels reported in Fig.~\ref{fgr:theory}, which we use to fit the Rabi and Ramsey data reported in this work. In Fig.~\ref{fgr:alldata}(b) we plot in teal the difference in energy between the ground and excited states; we plot in yellow the difference between these same states and the first excited state energy ($E_{1}/h$ = \amount{0.75}{GHz}). Although we do not directly observe this state, $E_{1}$, its presence is motivated by the FCI calculations described above, and we infer its existence and energy for reasons discussed below and in Sec.~S6 of Ref.~\cite{Suppl}. In the experiments, this state has nonzero initialization occupation both because of non-adiabaticity of the pulse sequence (Sec.~S7 in \cite{Suppl}) and because of thermal excitation caused by electron temperatures of about \amount{100}{ mK} ($k_{B}T$ = \amount{2.1}{GHz}).

The data shown in Fig.~\ref{fgr:alldata}(b) correspond to all the Rabi and Ramsey spectra reported in this work, as described in the legend. We plot the spectra from Fig.~\ref{fgr:SEM}(h,j) as purple circles and green triangles, and we fit to them the transitions $E_{05}$ and $E_{15}$, corresponding to the differences between the ground and first excited states to the fifth excited state.  The resonant frequencies from Fig.~\ref{fgr:manyres}(a-c) are shown as navy blue diamonds (for the resonance that moves up between Fig.~2(a) and Fig.~2(c)) and pink diamonds (for the resonance that moves down between Fig.~2(a) and Fig.~2(c)). As discussed earlier, these points merge with decreasing $f_{\mathrm{R}}$, providing additional evidence that these Rabi oscillations are driven from the ground and first excited state; if these two resonances belonged to the same dispersion, they would merge into a single chevron at the dispersion minimum instead of overlapping, as observed in Fig.~\ref{fgr:manyres}(c) and Fig.~\ref{fgr:additionaldoubledata} of Ref.~\cite{Suppl}. If both transitions occurred as excitations from the ground state, a level crossing would only occur if one of the tunnel couplings was anomalously low ($\leq$ 0.1~GHz) which is not supported by the shape of the Ramsey spectra. Finally, the light blue squares in Fig.~\ref{fgr:alldata}(b) show energies corresponding to Rabi oscillations reported in Fig.~\ref{fgr:allrabipoints} of Ref.~\cite{Suppl}. The density of transitions in frequency space 

\noindent as compared with the FCI calculations support the necessity to consider both $E_{0}$ and $E_{1}$ transitions. In total, Fig~\ref{fgr:alldata} summarizes the coherent control of eight different resonances in this Wigner molecule. 

For the quantum dot hybrid qubit, it is useful to be able to tune the singlet-triplet splitting, where the energies of interest are bounded by temperature on the low end and ease of microwave engineering on the high end. For silicon dots, the non-interacting singlet-triplet energy is given by the valley splitting, which depends on lateral confinement only through wavefunction overlap with interface roughness \cite{Friesen:2006p202106}. For strongly interacting electrons, this valley-like eigenstate gains contributions from many orbital levels, each of which depends strongly on lateral confinement. As we show in Fig.~\ref{fgr:eeinteraction}(b) of Ref.~\cite{Suppl}, the singlet-triplet splitting becomes strongly tunable in the presence of interactions. For the parameters considered here, varying the lateral confinement strength from 55~GHz to 90~GHz in the presence of interactions results in singlet-triplet splittings ranging from 0.18~GHz to 4.42~GHz, a variation of more than a factor of 20; in the non-interacting case this energy varies by only a factor of 2.6.

In conclusion,
we have demonstrated coherent manipulation of a dense ladder of Wigner molecule energy levels.
Eight transitions are controlled coherently using Rabi pulse sequences, and two levels are explored in more detail using Ramsey pulse sequences. A six-level model motivated by both experiment and FCI calculations is presented. The presence of such a Wigner-molecular regime with $E_{\mathrm{ee}} \gg E_{\mathrm{orb}}$ provides an additional tool for controlling the energy splittings in gate-defined quantum dots, enabling small changes in confinement to have a large impact on the energy splitting between adjacent levels.

This research was sponsored in part by the Army Research Office (ARO), through Grant No. W911NF-17-1-0274, and by the Vannevar Bush Faculty Fellowship program under ONR grant number N00014-15-1-0029.  JC acknowledges support from the National Science Foundation Graduate Research Fellowship Program under Grant No.\ DGE-1747503 and the Graduate School and the Office of the Vice Chancellor for Research and
Graduate Education at the University of Wisconsin-Madison with funding from the Wisconsin Alumni Research
Foundation. We acknowledge the use of facilities supported by the NSF through the UW-Madison MRSEC (DMR-1720415) and by the NSF MRI program (DMR-1625348). The views and conclusions contained in this paper are those of the authors and should not be interpreted as representing the official policies, either expressed or implied, of the ARO, NSF, or the U.S. Government. The U.S. Government is authorized to reproduce and distribute reprints for Government purposes notwithstanding any copyright notation herein.

\clearpage
\newpage
\mbox{~}
\onecolumngrid

\section{Supplementary material for `Coherent control and spectroscopy of a semiconductor quantum dot Wigner molecule'}
\twocolumngrid
\beginsupplement

\subsection{\label{sec:datashift}S1: Additional information for Figure 1}
\begin{figure}[H]
\centering
  \includegraphics[trim = 0 360 0 75, clip, width = 18cm]{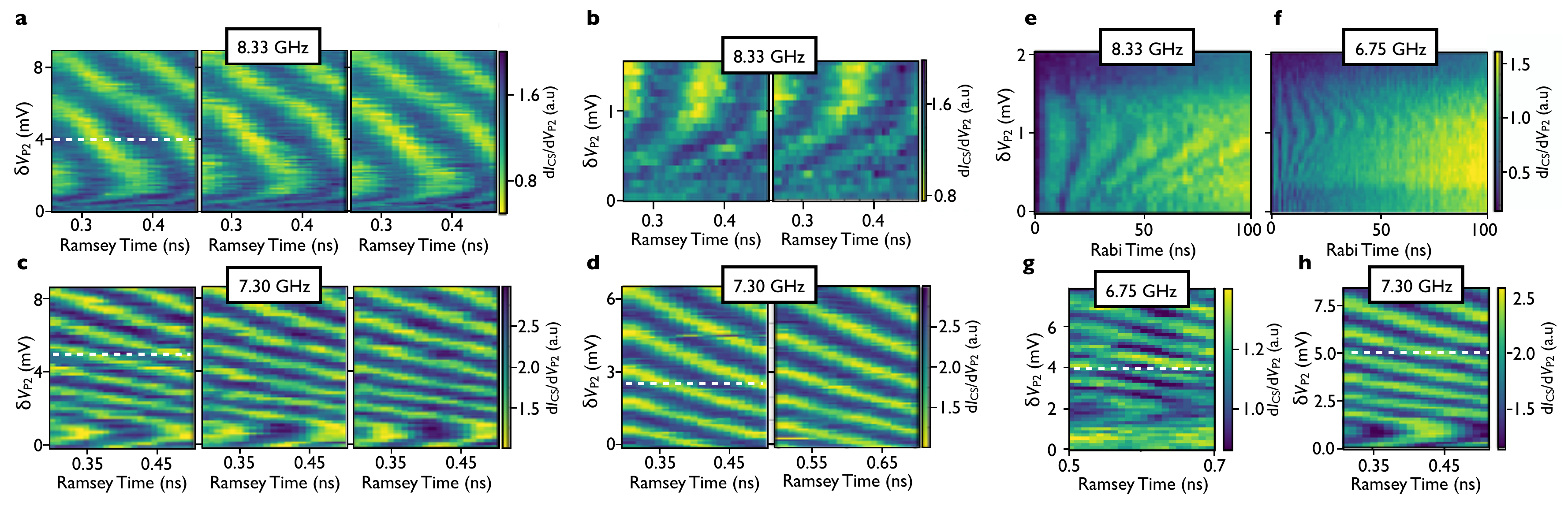}
  \caption{Ramsey oscillation data averaging and alignment. (a) Ramsey oscillations with $f_{\mathrm{R}} = \amount{8.33}{GHz}$ taken consecutively. These data are averaged together to produce Fig.~\ref{fgr:SEM}(h). (b) Ramsey oscillations with $f_{\mathrm{R}} = \amount{8.33}{GHz}$ taken consecutively. These data are averaged together, and the resulting oscillations, in addition to those from (a), are fit to produce the purple circles in Fig.~\ref{fgr:alldata}(b). (c) Ramsey oscillations with $f_{\mathrm{R}} = \amount{7.30}{GHz}$ taken consecutively. These data are averaged together to produce Fig.~\ref{fgr:SEM}(j). (d) Ramsey oscillations with $f_{\mathrm{R}} = \amount{7.30}{GHz}$ taken consecutively. The oscillations in these data are fit and the corresponding fits are averaged together. Those points, in addition to those resulting from fitting (c),  produce the green triangles in Fig.~\ref{fgr:alldata}(b). (e) Rabi oscillations for the same resonance as Fig.~\ref{fgr:SEM}(g), with $f_{\mathrm{R}} = \amount{8.33}{GHz}$. (f) Rabi oscillations for the same energy level as Fig.~\ref{fgr:SEM}(i) but a different detuning, now with  $f_{\mathrm{R}} = \amount{6.75}{GHz}$. This data set is taken directly after (e), with the only change being the microwave drive. The centers of the two chevrons both correspond to $\delta V_{\mathrm{P2}} = \amount{1.1}{mV}$. (g) Ramsey fringes corresponding to the resonance in (f) taken at $f_{\mathrm{R}} = \amount{6.75}{GHz}$. (h) Ramsey fringes,taken at a different detuning, with the same $f_{\mathrm{R}} = \amount{7.3}{GHz}$ as Fig.~\ref{fgr:SEM}(j). With the dispersion minima as reference, we use frequency data extracted from (h), and shift it accordingly so it reflects the detuning location in (g). This allows us to plot in Fig.~\ref{fgr:alldata}(b) the coexistance of both dispersions. The white dashed lines in (a,c,d,g,h) denote the value of $\delta V_{\mathrm{P2}}$ for which the Ramsey detuning is zero.
}

  \label{fgr:datashift}
\end{figure}

The data in Fig.~\ref{fgr:SEM} presents two distinct dispersions that are individually addressable. To produce  Fig.~\ref{fgr:SEM}(h), we averaged three data sets taken consecutively, as shown in Fig.~\ref{fgr:datashift}(a). 
Additional data at low detuning is seen in Fig.~\ref{fgr:datashift}(b). 
The purple circles in Fig.~\ref{fgr:alldata}(b) are produced by fitting the resulting data from Fig.~\ref{fgr:datashift}(a,b) for the frequency of oscillation. Similarly, to produce Fig.~\ref{fgr:SEM}(j) we averaged the data shown in Fig.~\ref{fgr:datashift}(c). Additional data taken at high detuning is shown in Fig.~\ref{fgr:datashift}(d). Since the two plots in Fig.~\ref{fgr:datashift}(d) are at different time steps, these data are first fit for their oscillation frequencies and the two fits are averaged together. The green triangles in Fig.~\ref{fgr:alldata}(b) are produced by fitting the results from Fig.~\ref{fgr:datashift}(c,d).

The differences in size of the Rabi oscillations in Fig.~\ref{fgr:SEM}(g,i) and the differences in shape between Fig.~\ref{fgr:SEM}(g,i) and Fig.~\ref{fgr:SEM}(h,j) for the two microwave frequencies are indicative of two distinct energy levels. In order to map the resulting dispersions from Fig.~\ref{fgr:SEM}(h,j) relative to each other, Fig.~\ref{fgr:SEM}(g,i) are repeated at the same detuning, achieved using the 
microwave drives of  $f_{\mathrm{R}}$ = \amount{8.33}{GHz} and $f_{\mathrm{R}}$ = \amount{6.75}{GHz}, as shown in Fig.~\ref{fgr:datashift}(e,f). Rabi frequencies extracted from Fig.~\ref{fgr:datashift}(e,f) are represented in Fig.~\ref{fgr:alldata}(b) by including the $f_{\mathrm{R}}$ = \amount{6.75}{GHz} magenta star at the same detuning as the $f_{\mathrm{R}}$ = \amount{8.33}{GHz} orange star. 
Though we do not observe off resonant oscillations from one spectrum in measurements of the other, it does appear that the region of suppressed coherence in Fig.~\ref{fgr:datashift}(e) is the region of enhanced coherence in Fig.~\ref{fgr:datashift}(f) and visa versa. 

The resonance in Fig.~\ref{fgr:datashift}(f) belongs to the same energy level as that in Fig.~\ref{fgr:SEM}(i), as can be seen by comparing Ramsey spectroscopy taken at the two $f_{\mathrm{R}}$ locations, as shown in Fig.~\ref{fgr:datashift}(g,h). The general shape of the data in Fig.~\ref{fgr:datashift}(g,h) is consistent, and the only difference is a relative shift in $\delta V_{\mathrm{P2}}$ by \amount{0.96}{mV}. To produce the points in Fig.~\ref{fgr:alldata}(b), we shifted the data in Fig.~\ref{fgr:datashift}(d,h) by $\delta V_{\mathrm{P2}}$ = \amount{0.96}{mV} so that they match the detuning in Fig.~\ref{fgr:datashift}(g),

\onecolumngrid

 \begin{figure}[H]
\centering
  \includegraphics[trim = 82 595 730 25, clip, width = 17.8cm]{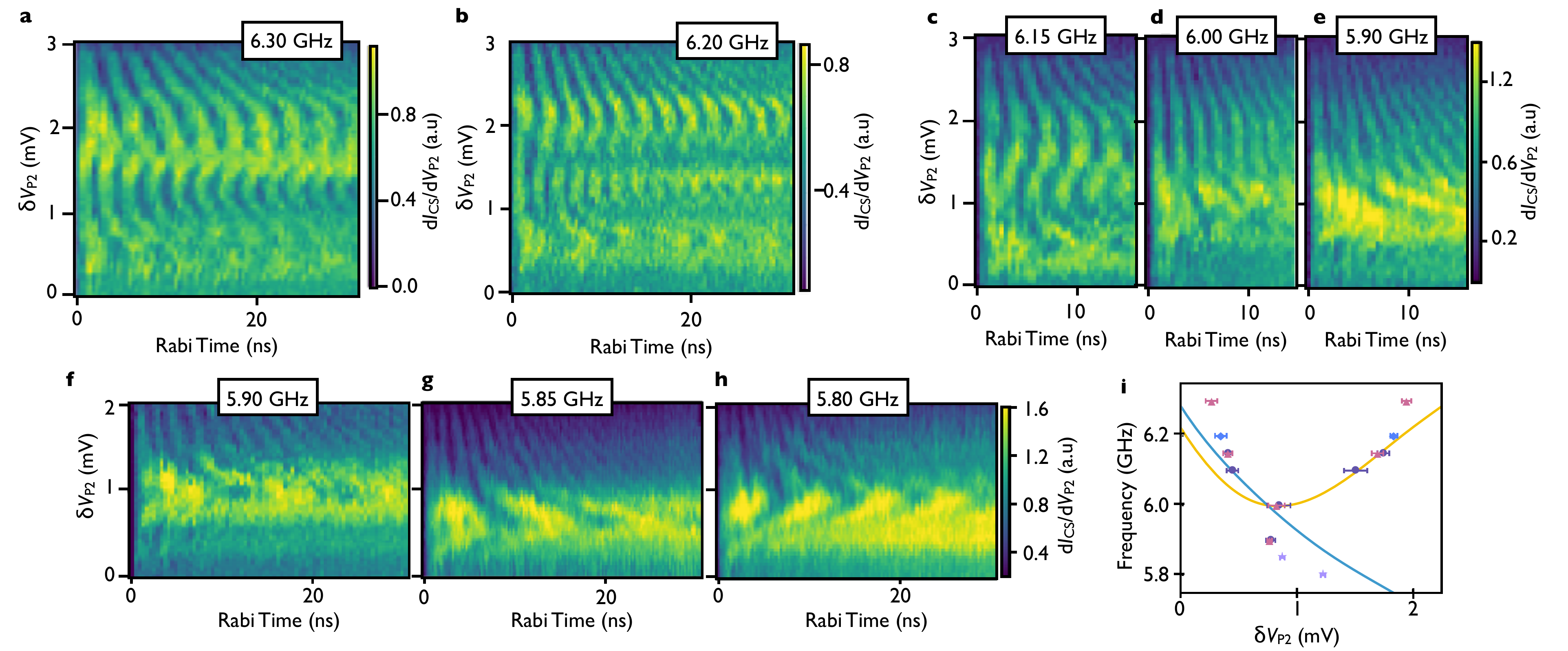}
  \caption{Additional level crossing data corresponding to the dark blue diamonds in Fig~\ref{fgr:alldata}(b). (a,b) Rabi oscillations at $f_{\mathrm{R}}$ = \amount{6.30}{GHz} and $f_{\mathrm{R}}$ = \amount{6.20}{GHz} . (c, d)Additional Rabi oscillations at $f_{\mathrm{R}}$ = \amount{6.15}{GHz} and $f_{\mathrm{R}}$ = \amount{6.00}{GHz} used to combine multiple sets of data with different $V_{P2}$ offsets. (e,f) Rabi oscillations with  $f_{\mathrm{R}}$ = \amount{5.90}{GHz}, again used for combining multiple data sets. (g) Rabi oscillations with  $f_{\mathrm{R}}$ = \amount{5.85}{GHz}. (h) Rabi oscillations with $f_{\mathrm{R}}$ = \amount{5.80}{GHz}. (i) All of the data from the navy and pink diamonds in Fig~\ref{fgr:alldata}(b). Different markers correspond to data taken consecutively before device drift, as described in Table \ref{tab:levelcrossing}. Blue and orange lines correspond to the interpolation functions $f(\varepsilon)$ and $g(\varepsilon)$, respectfully, which are described in section S2.}

  \label{fgr:additionaldoubledata}
\end{figure}
\twocolumngrid

\noindent  as experimentally measured. The data in Fig.~\ref{fgr:alldata}(b) is converted from gate voltage to detuning with a detuning alpha of $\alpha_{\varepsilon, P2} = 0.085 \pm 0.009$. 

\subsection{\label{sec:doubleresdata}S2: Additional information on data and simulation for Figure 2}
\begin{table}[htp]
\centering
\setlength{\tabcolsep}{8pt}
\caption{\label{tab:dataPanels}%
Sets of on-resonance oscillations discussed in Fig.~\ref{fgr:manyres} and Fig.~\ref{fgr:additionaldoubledata}. Each set is separated by a line.} 
\begin{tabular}{ccccc}
\hline
\hline
\multicolumn{2}{c}{Set1}&&\multicolumn{2}{c}{Set2}\\
\cline{1-2} \cline{4-5}
$f_{\mathrm{R}}$ (GHz) & Figure && $f_{\mathrm{R}}$ (GHz) & Figure \\
\hline
 6.30 & Fig.\ref{fgr:additionaldoubledata}(a) && 6.15 & Fig.\ref{fgr:manyres}(a) \\
 6.15 & Fig.\ref{fgr:additionaldoubledata}(c) && 6.10 & Fig.\ref{fgr:manyres}(b) \\
 6.00 & Fig.\ref{fgr:additionaldoubledata}(d) && 6.00 & Fig.\ref{fgr:manyres}(c\\
 5.90 & Fig.\ref{fgr:additionaldoubledata}(e) && 5.90 & Fig.\ref{fgr:additionaldoubledata}(f)\\
 \\
\hline
\multicolumn{2}{c}{Set3}&&\multicolumn{2}{c}{Set4} \\
\cline{1-2} \cline{4-5}
$f_{\mathrm{R}}$ (GHz) & Figure && $f_{\mathrm{R}}$ (GHz) & Figure \\
\hline
5.85 & Fig.\ref{fgr:additionaldoubledata}(g)&&6.20& Fig.\ref{fgr:additionaldoubledata}(b)\\
5.80 &  Fig.\ref{fgr:additionaldoubledata}(h)&& - \\

\hline
\hline
\end{tabular}

\end{table}

Figures \ref{fgr:manyres} and  \ref{fgr:additionaldoubledata} contain all the data used to extract the navy blue and pink diamonds in Fig.~\ref{fgr:alldata}(b). Each diamond in Fig.~\ref{fgr:alldata}(b) represents the center of a Rabi oscillation for a given drive frequency. As seen in Fig.~\ref{fgr:manyres}, \ref{fgr:additionaldoubledata}, many of the data sets have two clear on-resonance oscillations. In these cases, both centers are plotted with the same frequency and different value of $\delta V_{\mathrm{P2}}$. All of the extracted data from these two figures can be seen in  Fig.~\ref{fgr:additionaldoubledata}(i), where the different markers correspond to different sets of data taken consecutively (before the device shifted). These sets are enumerated in Table \ref{tab:dataPanels}. The uncertainty in Fig.~\ref{fgr:additionaldoubledata}(i) is the uncertainty in the center of the oscillation, while the uncertainty in these points in Fig.~\ref{fgr:alldata}(b) includes uncertainty in their location relative to zero double-dot detuning. 

The Rabi measurements shown in Fig.~\ref{fgr:manyres} indicate multi-level structure. We find that the best model that reproduces the main features of the experimental measurements in a simple way consists of four relevant states: two partially populated lower levels, separated by a small energy $E_{01}$, that make transitions to two excited states under detuning driving fields. The effective Hamiltonian of this four-level system is
\begin{equation}
\label{eq:JCH0}
H_0^{\rm toy}=\begin{pmatrix}
0 & 0 & 0 & 0 \\
0 & E_{01} & 0 & 0 \\
0 & 0 & f(\varepsilon) & 0 \\
0 & 0 & 0 & g(\varepsilon)+E_{01}
\end{pmatrix},
\end{equation}

\noindent where $f(\varepsilon)$ and $g(\varepsilon)$ are interpolation functions based on the energy measurements, shown as blue and orange lines in Fig. \ref{fgr:additionaldoubledata}(i). We set $E_{01}/h=3$ GHz, which is compatible with the small energy gap between the lowest two states observed in the experiment and large enough to avoid undesired interactions. The driving field is characterized by
\begin{equation}
\label{eq:JCH1}
H_1^{\rm toy}=\delta\varepsilon\cos(\omega t)\begin{pmatrix}
0 & 0 & R_{02} & 0 \\
0 & 0 & 0 & R_{13} \\
R_{02} & 0 & 0 & 0 \\
0 & R_{13} & 0 & 0
\end{pmatrix}
\end{equation}

where $R_{02}$ and $R_{13}$ are fitting parameters. The only other fitting parameter is the ground state population 

\onecolumngrid

\begin{figure}[H]
\centering
\includegraphics[trim = 30 480 280 30, clip, width = 18.0cm]{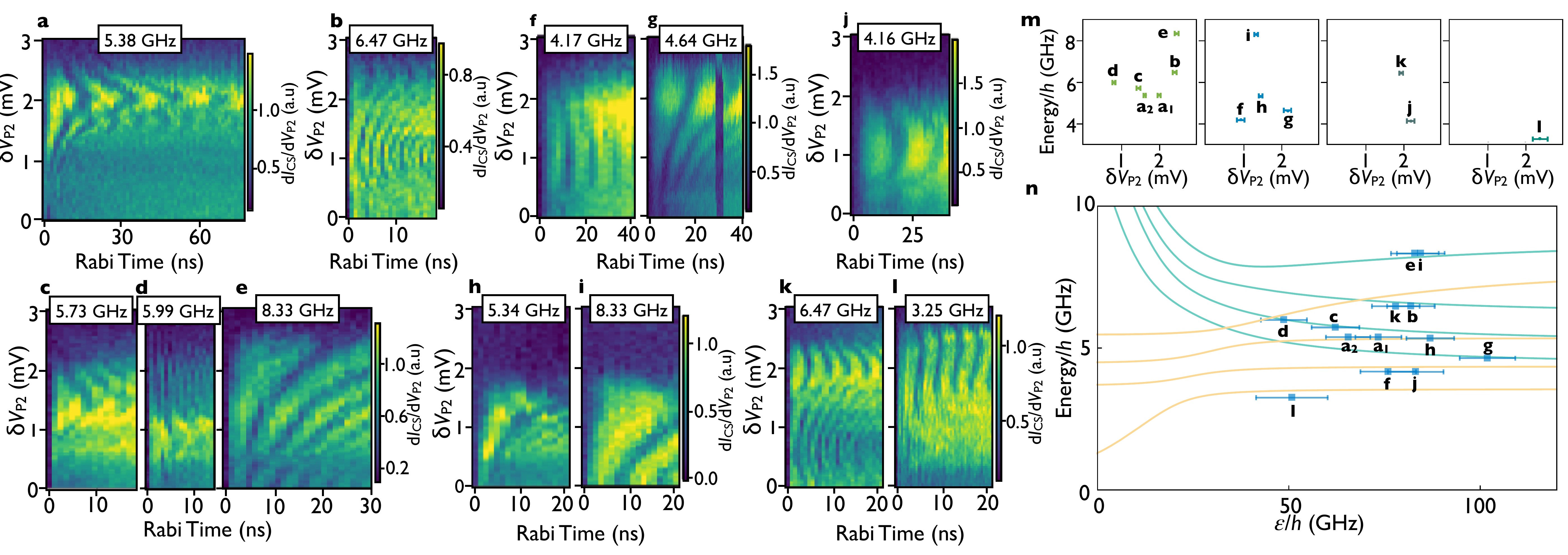}
  \caption{
 Rabi oscillations associated with the blue squares in Fig~\ref{fgr:alldata}(b), with frequencies (a) $f_{\mathrm{R}}$ = \amount{5.38}{GHz}, (b) $f_{\mathrm{R}}$ = \amount{6.47}{GHz}, (c) $f_{\mathrm{R}}$ = \amount{5.73}{GHz}, (d) $f_{\mathrm{R}}$ = \amount{5.99}{GHz}, (e) $f_{\mathrm{R}}$ = \amount{8.33}{GHz}, (f) $f_{\mathrm{R}}$ = \amount{4.17}{GHz}, (g) $f_{\mathrm{R}}$ = \amount{4.64}{GHz}, (h) $f_{\mathrm{R}}$ = \amount{5.34}{GHz}, (i) $f_{\mathrm{R}}$ = \amount{8.33}{GHz}, (j) $f_{\mathrm{R}}$ = \amount{4.16}{GHz}, (k) $f_{\mathrm{R}}$ = \amount{6.47}{GHz}, (l) $\delta V_{\mathrm{P2}}$= \amount{3.25}{GHz}. (m) (a-l) were taken in four sets, where within each set there is no device shifting and corresponding adjustment. The sets are (a-e), (f-i), (j,k), and (l). The points in (m) represent the location in relative $P_{2}$ of the center of the Rabi oscillation for a given driving frequency. (n) These four sets are then placed onto the model by simultaneously matching (e,i), and (b,k), as well as looking at the distance of (e,l) from the approximate polarization line.}. 
  \label{fgr:allrabipoints}
\end{figure}

\twocolumngrid 

\begin{figure}[h!]
\centering
  \includegraphics[trim = 25 540 805 120, clip, width = 6cm]{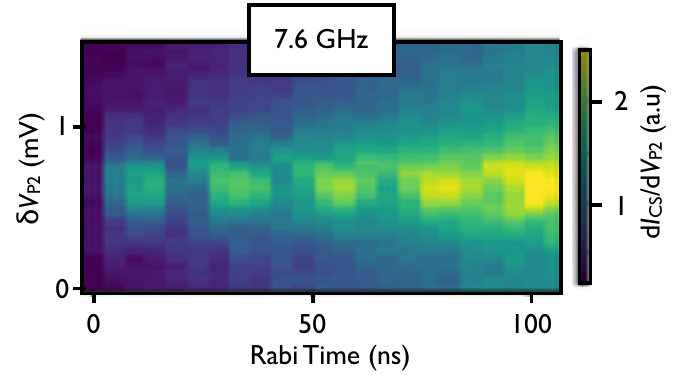}
  \caption{
 Rabi oscillations associated $f_{\mathrm{R}}$ = \amount{7.60}{GHz}, shown as a light purple star in Fig.~\ref{fgr:alldata}(b).}

  \label{fgr:rabi76}
\end{figure}

\begin{table}[H]
\centering
\caption{\label{tab:levelcrossing}%
Parameters $R_{02}$ and $R_{13}$ from Eq.~\ref{eq:JCH1}, in addition to the drive frequency $f_{\mathrm{R}}$ and ground state population $\rho_{00}$, which are all used for the simulations in Fig.~\ref{fgr:manyres}(d-f).} 
\setlength{\tabcolsep}{10pt}
\begin{tabular}{cccc}
\hline
\hline
&Fig.\ref{fgr:manyres}(d)&Fig.\ref{fgr:manyres}(e)&Fig.\ref{fgr:manyres}(f)\\
\hline
$f_{\mathrm{R}}$ (GHz) & 6.15  & 6.1  & 6.0 \tabularnewline
$R_{02}\times 10^3$ & 0.443  & 0.463  & 0.616 \tabularnewline

$R_{13}\times 10^3$ & 0.925 & 0.984 & 1.626 \tabularnewline

$\rho_{00}$ & 0.7 & 0.7 & 0.65 \tabularnewline
\hline
\hline
\end{tabular}
\end{table}

\noindent $\rho_{00}$, assuming $\rho_{11}=1-\rho_{00}$. The fitting parameters used for each case are shown in Table \ref{tab:levelcrossing}. 

We find that an even simpler model, such as a three-level system with a singly-populated ground state and two interacting excited states, might also be an explanation. However, the simulations with this simpler model added more unobserved interference phenomena. We note that the simple four-level model is rather limited since the values of the Rabi parameters $R_{02}$ and $R_{13}$ should depend on detuning. Moreover, the energy level landscape is more complicated than a four-level system. Nevertheless, this simple model captures the main features in the measurements.

Finally, to simulate the presence of noise in Fig. 3, we perform a Gaussian convolution with $\sigma_\varepsilon=7.1\mu$eV. This number is extracted by averaging the $\sigma_\varepsilon =\frac{ \sqrt{2} \hbar} {\vert \frac{dE}{d\varepsilon}\vert T_{2}^{*}}$ computed with $T_{2}^{*}$ and $\frac{dE}{d\varepsilon}$ from the \amount{7.30}{GHz} and \amount{8.33}{GHz} resonances shown in Fig.~\ref{fgr:manyres}.

\subsection{\label{sec:7resdata}S3: Additional data for Figure 4}
Figure \ref{fgr:allrabipoints}, \ref{fgr:rabi76} contains the measured Rabi oscillations corresponding to the plots in Fig~\ref{fgr:alldata}(b).  The data in Fig.~\ref{fgr:allrabipoints} is grouped in four sets, (a-e), (f-i), (j,k), and (l), as shown in (m). Each set represents a portion of data that is taken without any drift of the DQD device, and without changing any pulse-shape parameters besides microwave frequency. 
Each extracted point is obtained by looking at line-cuts in time through the data and estimating the center point of the oscillation in gate voltage. Note that all the data in these figures span \amount{3}{mV} on the gate which dictates the quantum dot detuning, $P_{2}$. The drastic changes in size, shape, and frequency of these oscillations in the same relative window of detuning supports the interpretation that these oscillations correspond to different energy level transitions. The data from (m) is mapped to the detuning axis in (n) by trying to simultaneously have both (e,i) and (b,k) be as close as possible. The location of (e,l) relative to the polarization line, which should be at the zero detuning point for the DQD, is approximated using their resonant locations within the latched readout region, stability diagrams and the pulse height.   

The data in Fig.~\ref{fgr:rabi76} appears as a purple star in Fig.~4(b). 

\subsection{\label{sec:sim}S4: Simulation details for Figure 3}
The simulation results shown in Fig.~\ref{fgr:theory} are performed by combining a full configuration interaction (FCI) approach with the empirical tight-binding (TB) theory of Boykin et al. \cite{Boykin:2004p115,Boykin:2004p165325}. In the TB model, the nearest-neighbor and the next nearest neighbor hopping parameters $t_1$ = 0.6829 and $t_2$ = 0.6119 are chosen such that the location of the resulting band minima coincide with the location of bulk silicon's conduction band minima and the effective mass, determined by the curvature of the conduction band at its minima, is equal to the lateral effective mass of bulk silicon. With an additional hopping parameter $t_3$, chosen such that the transversal effective mass of bulk Si is obtained, the model becomes two-dimensional in the x-z plane. We are thus able to study the effects of different interface profiles by simply assigning on-site terms $E_{Si}$ and $E_{SiGe}$ to grid points in different patterns. Resulting wave functions have fast oscillations in z that are responsible for the breaking of valley degeneracy,  and disorder at the interface causes valley and orbital degrees of freedom to couple. In the calculations, we treat the third dimension y analytically with the assumption that the confinement potential is parabolic. 

While TB provides an accurate description of single electron wave functions by capturing the valley physics of silicon and allowing modeling of the quantum well interface disorder, FCI allows us to calculate two electron energies by including the effects of electron-electron interactions. After adding the spin degree of freedom, we generate all possible two-electron Slater determinants based on the 45 lowest energy TB eigenstates, which constitute the basis for the FCI calculation.  We calculate the full Hamiltonian, including the electron-electron interaction term (with dielectric constant 11.4 \cite{Faulkner:1969p713}), in this basis and diagonalize it to obtain the two-electron energy eigenvalues and eigenstates. For the simulation in Fig.~\ref{fgr:theory}, we used a tilted quantum well interface with a tilting angle of $\sim 0.2^{\circ}$, quantum well width of \amount{9.1}{nm} and an electric field of \amount{0.6}{MV/m} perpendicular to the interface. With these parameters, valley splittings range from 2.4 to 4.8~GHz (10 to \amount{20}{\mu eV}) in the considered $\hbar \omega_x$ domain, and in particular is 3.81~GHz (\amount{14}{\mu eV}) for the $\omega_x/2\pi$ = 59.2 GHz used in Fig.~\ref{fgr:theory}(a). We pick the location of the center of the dot with respect to the uniform steps at the interface to obtain the best agreement with the experimental data.

\subsection{S5: Electron-electron interactions}

\begin{figure}[h]
\centering
  \includegraphics[trim = 35 310 720 30, clip, width = 8cm]{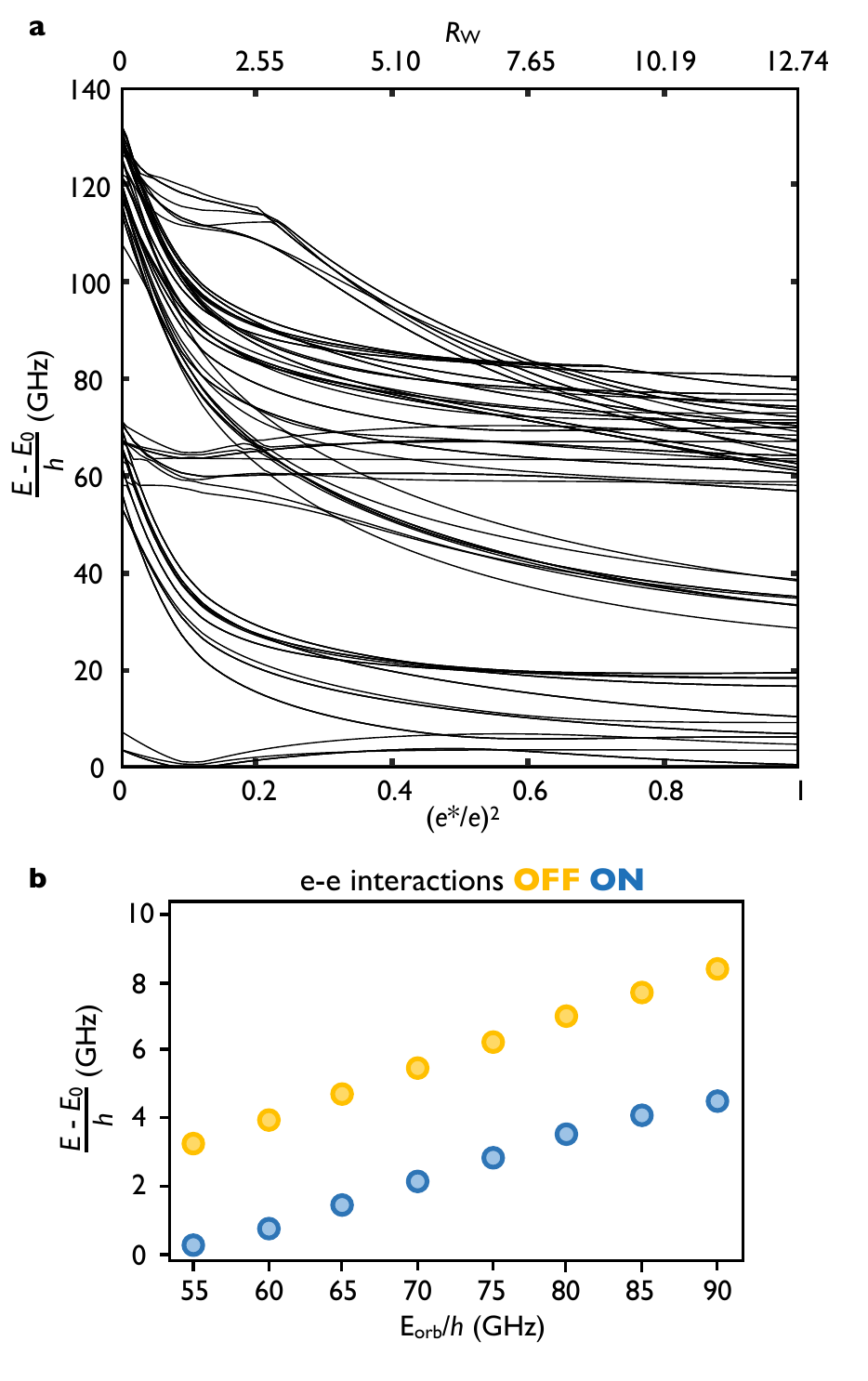}
  \caption{
  (a) Two-electron energy level separation from the ground state at $E_{\mathrm{orb}}/h$ = 59.2 GHz plotted as a function of the relative interaction strength $(e^{*}/e)^{2}$ (bottom) and Wigner parameter $R_{\mathrm{W}}$ (top), showing suppression of excitation energies due to electron-electron interactions. (b) Two-electron energy levels with electron-electron interactions turned off (on) are plotted in yellow (blue) as a function of orbital confinement.}

  \label{fgr:eeinteraction}
\end{figure}

As described in the main text, the Wigner parameter $R_{\mathrm{W}}$ = $E_{\mathrm{ee}}$ / $E_{\mathrm{orb}}$ is used to classify Wigner molecules. Here we take $E_{\mathrm{orb}} = \hbar\omega_{x}$ and calculate the Coulomb energy of two point charges separated by the characteristic length scale of the quantum dot, which is a typical measure of the electronic repulsion in the system. This  results in $E_{\mathrm{ee}} = \frac{e^2 \sqrt{m_{t} \omega_{x} / \hbar}}{4 \pi \epsilon}$, where $m_{t}$ is the effective mass. Using the $m_{t}$ and relative permittivity in Si, with $E_{\mathrm{orb}}/h$ = 59.2 GHz consistent with the FCI calculations presented in Fig.~\ref{fgr:theory}, we obtain $R_{\mathrm{W}} = 12.74$. 

To closely observe the effects of electron-electron interactions on the energy spectrum, we gradually introduce them in our simulations by using an artificial electron charge e* that ranges from 0 to real electron charge e. Two-electron energy level splittings from the ground state are plotted in Fig.~\ref{fgr:eeinteraction} as a function of both $(e^{*}/e)^{2}$ and $R_\mathrm{W}$. In the absence of electron-electron interactions, the two-electron energy spectrum is dictated by smaller (5-15 GHz) valley splittings and larger (60 GHz) orbital splittings. With increasing interaction strength, two-electron wave functions are formed with contributions from many excited valley and orbital states, as the electrons are increasingly driven away from each other. This hybridization results in densely packed energy levels in the strong interaction regime, which is seen in Fig.~\ref{fgr:eeinteraction}. 

\subsection{\label{sec:hamparams}S6: Choice of Hamiltonian Parameters for Figure 4}
\begin{table}[h]
\centering
\setlength{\tabcolsep}{6pt}
\addtolength{\parskip}{-.1mm}
\caption{\label{tab:HamParam}%
Parameters for Eq.~\ref{eq:Hamiltonian} chosen to best fit the data and percent change in each parameter that degrades the fit.} 
\begin{tabular}{ccccccc}
\hline
\hline
 \multicolumn{3}{c}{Energy}&&  \multicolumn{3}{c}{Tunnel Coupling}\\
 \cline{1-3} \cline{5-7}
 $E_{n}$ & E/$h$ (GHz)& $\delta E$ (\%) && $\Delta_{n}$ & $\Delta/h$ (GHz)& $\delta \Delta$ (\%)\\
 \hline 
- & - & - && $\Delta_{0}$ & 6 & 10 \\
$E_{1}$& 0.75 & 20 && $\Delta_{1}$ & 1.5 & 30\\
$E_{2}$ & 4.3 & 10 && $\Delta_{2}$ & 1.5 & 30\\
$E_{3}$ & 5.1 & 8 && $\Delta_{3}$ & 1.5 & 30 \\
$E_{4}$ & 6.1 & 5 && $\Delta_{4}$ &1.5 & 30\\
$E_{5}$ & 9 & 4 && $\Delta_{5}$ & 10.5 & 10\\
\hline
\hline
\\

\end{tabular}

\end{table}

The lines in Fig.~\ref{fgr:alldata}(a) are the eigenvalues of 
\begin{equation}
\label{eq:Hamiltonian}
H=\begin{pmatrix}
 \varepsilon /2 & \Delta_{0} & \Delta_{1} & \hdots &  \Delta_{5} \\
\Delta_{0} & -\varepsilon /2 & 0 & \hdots & 0 \\
\Delta_{1} & 0 & -\varepsilon /2 + E_{1} & \hdots & 0 \\
\Delta_{2} & 0 & 0 & \ddots & 0 \\
\Delta_{3} & 0 & 0 &  \hdots&0 \\
\Delta_{4} & 0 & 0 &  \hdots & 0 \\
\Delta_{5} & 0 & 0 &  \hdots& -\varepsilon /2 + E_{5} \\
\end{pmatrix},
\end{equation}
which describes a system of five excited states in the right quantum dot coupled to the left quantum dot. The values for energies $E_{n}$ and tunnel couplings $\Delta_{n}$ are listed in Table~\ref{tab:HamParam}. Though we have a large quantity of data, including two detuned-Ramsey measurements of different levels (Fig.~\ref{fgr:SEM}(h,j)), Rabi measurements of eight different levels(Fig.~\ref{fgr:SEM}(g,i), Fig.~\ref{fgr:allrabipoints}), and Rabi measurements of two merging levels(Fig.~\ref{fgr:manyres}(a-c), Fig.~\ref{fgr:additionaldoubledata}), the five-level model from Fig.~\ref{fgr:alldata}(b) based on Eq.~\ref{eq:Hamiltonian} is still under-constrained.  It has 11 independent parameters: five energies, $E_1$ -- $E_5$, and six tunnel couplings, $\Delta_0$ -- $\Delta_5$.

The detuned Ramsey data from Fig.~\ref{fgr:SEM}(h) (purple circles in Fig.~\ref{fgr:alldata}(b)) provides an approximate value for $E_{5}$ and $\Delta_{5}$, which respectively govern the high-detuning asymptote and low-detuning shape of $E_{05}$. Similarly, the set of detuned Ramsey data from Fig.~\ref{fgr:SEM}(j) (green triangles in Fig.~\ref{fgr:alldata}(b)) places strong constraints on $E_{15}$. Since $E_{05} - E_{15} = E_{01}$, these two Ramsey sets of data together constrain the high-detuning asymptote and low-detuning shape of $E_{01}$, thereby dictating the values of $E_{1}$ and $\Delta_{0}$. With this constraint on $E_{01}$, the relation $E_{0n} - E_{1n} = E_{01}$ narrows the values of the other energies $E_{2}$-$E_{4}$. Each energy $E_{n}$ has at least two Rabi measurements (one for $E_{0n}$ and one for $E_{1n}$) to define its value. Table~\ref{tab:HamParam} shows the percent change in each energy $E_{n}$ which degrades the fit. 

The individual tunnel couplings $\Delta_{1}-\Delta_{4}$ are not as firmly constrained as the energy values. However, as all the eigenvalues are interdependent, the individual tunnel couplings as well as the sum of the tunnel couplings are constrained to be within a certain range. For example, having any or all of $\Delta_{n}$ too large will push both $E_{01}$ downward and $E_{05}$ upward, both of which are well constrained, as discussed above. Table~\ref{tab:HamParam} shows the percent change in each tunnel coupling $\Delta_{0}$ and $\Delta_{5}$, as well as the percent change for all the tunnel couplings $\Delta_{1}$-$\Delta_{4}$ which are set equal, that degrades the fit.

\subsection{S7: Estimations of excited-state populations}
\begin{table}[H]
\centering
\setlength{\tabcolsep}{8pt}
\caption{\label{tab:populations}%
Expected ground ($\rho_{0}$) and excited state ($\rho_{1,3}$) populations based on the pulse sequence used.} 
\begin{tabular}{ccccc}
\hline
\hline

 & Pulse 1  & Pulse 2  & Pulse 3 & Pulse 4\tabularnewline
 \hline
$\varepsilon_h\ (\mu$eV) & 680  & 646  & 612 & 816 \tabularnewline

$t_g\ (ns)$ & 6  & 6  & 6 & 9 \tabularnewline

$\rho_{0}$ & 0.841  & 0.873  & 0.872 & 0.864 \tabularnewline

$\rho_{1}$ & 0.082 & 0.083 & 0.082 &  0.076 \tabularnewline

$\rho_{3}$ & 0.054 & 0.023 & 0.026 &  0.038 \tabularnewline
\hline
\hline
\end{tabular}
\end{table}

In the experiment, the qubit is initialized at very negative detuning values. In this region, the ground state is well-separated from the first-excited state, ensuring proper initialization. To manipulate the qubit, the detuning is pulsed from this region to positive detuning. Due to the large number of low-lying excited states, some excited states may be populated during this process. To estimate the excited-state populations, we simulate the initialization pulse to the positive detuned regime.

To describe the qubit, we consider the effective model given by Eq.~\ref{eq:Hamiltonian}. This model, while an approximation, provides intuition on the interactions with the excited states. In the experiment, the initial detuning value ranges from $-400$ to \amount{-300}{\mu eV} ($-97$ to \amount{-73}{GHz}). Since the experimental pulses go from the negative far-detuned regime to the positive detuned regime, the exact initial detuning position is not relevant; hence, we choose $\varepsilon_0=-350\mu$eV. The simulated pulse is simply $\varepsilon(t)=\varepsilon_0+\varepsilon_h/t_g\cdot t$, where $\varepsilon_h$ is the pulse height and $t_g$ the duration of the pulse. 

We consider the four different experimental pulses in Table \ref{tab:populations}, which are representative of the pulses used in this work. The three states with the highest initialization population are also shown in the table. The results show that the ground state is the most populated, and that there is non-negligible population in the first excited state, consistent with the experimental results reported in the main text. The third excited state, which has the next highest initialization population, is less populated and not observed in the experiments.

\subsection{\label{sec:pulse correction}S8: Pulses and pulse corrections}
Pulses are generated using a two-channel Textronix AWG 70002A and changing the relative skew between the two channels. We perform lock-in measurements at the frequency with which we modulate on and off the application of microwave pulses. Each pulse consists of a short manipulation phase followed by a longer measurement phase of typical duration \amount{1.2}{\mu s}. The typical modulation frequency is around 10kHz. The reported Ramsey fringes are the amplitude of the lock-in signal

The pulse sequences we use consist of quasi-dc changes in detuning combined with ac microwave bursts. Because of the frequency dependent attenuation down the dilution refrigerator, we apply pulse corrections to make our dc pulses more similar to the intended shape. We use S21 measurements of the high frequency lines to determine the frequency dependent attenuation expected. We then take a FFT for a step function, and weight the spectrum according to our S21 measurements. An inverse FFT is taken to obtain the corrected pulse segment, which becomes the building block for our new pulse sequence. We apply the corrected building block for every change in dc level to our intended dc pulse sequence. This pulse sequence visually looks like an overcorrection of the pulse in order to counteract the measured finite rise-time. We finally add in any ac components to our pulses. After these pulse corrections, there is still an observable drift in the top of the pulse as a function of time for short times (Ramsey measurements). To place points on Fig.~~\ref{fgr:alldata}(b) it is important to know the detuning. Thus, we perform additional post-processing to the Ramsey data by shifting line scans with the function $A (1+ b e^{-t/\tau })$ in order to align the dispersion minima of consecutive Ramsey oscillations. This allows us to accurately plot detuned Ramsey data for small times, such as the purple circles and green triangles in Fig.~\ref{fgr:alldata}(b). Here, A is the detuning pulse amplitude, $b=0.2$, and $\tau$ = \amount{1.8}{ns}. 

\section{\label{sec:citeref}References}
\bibliography{main}

\begin{thebibliography}{37}%
\makeatletter
\providecommand \@ifxundefined [1]{%
 \@ifx{#1\undefined}
}%
\providecommand \@ifnum [1]{%
 \ifnum #1\expandafter \@firstoftwo
 \else \expandafter \@secondoftwo
 \fi
}%
\providecommand \@ifx [1]{%
 \ifx #1\expandafter \@firstoftwo
 \else \expandafter \@secondoftwo
 \fi
}%
\providecommand \natexlab [1]{#1}%
\providecommand \enquote  [1]{``#1''}%
\providecommand \bibnamefont  [1]{#1}%
\providecommand \bibfnamefont [1]{#1}%
\providecommand \citenamefont [1]{#1}%
\providecommand \href@noop [0]{\@secondoftwo}%
\providecommand \href [0]{\begingroup \@sanitize@url \@href}%
\providecommand \@href[1]{\@@startlink{#1}\@@href}%
\providecommand \@@href[1]{\endgroup#1\@@endlink}%
\providecommand \@sanitize@url [0]{\catcode `\\12\catcode `\$12\catcode
  `\&12\catcode `\#12\catcode `\^12\catcode `\_12\catcode `\%12\relax}%
\providecommand \@@startlink[1]{}%
\providecommand \@@endlink[0]{}%
\providecommand \url  [0]{\begingroup\@sanitize@url \@url }%
\providecommand \@url [1]{\endgroup\@href {#1}{\urlprefix }}%
\providecommand \urlprefix  [0]{URL }%
\providecommand \Eprint [0]{\href }%
\providecommand \doibase [0]{http://dx.doi.org/}%
\providecommand \selectlanguage [0]{\@gobble}%
\providecommand \bibinfo  [0]{\@secondoftwo}%
\providecommand \bibfield  [0]{\@secondoftwo}%
\providecommand \translation [1]{[#1]}%
\providecommand \BibitemOpen [0]{}%
\providecommand \bibitemStop [0]{}%
\providecommand \bibitemNoStop [0]{.\EOS\space}%
\providecommand \EOS [0]{\spacefactor3000\relax}%
\providecommand \BibitemShut  [1]{\csname bibitem#1\endcsname}%
\let\auto@bib@innerbib\@empty
\bibitem [{\citenamefont {Yang}\ \emph {et~al.}(2020)\citenamefont {Yang},
  \citenamefont {Leon}, \citenamefont {Hwang}, \citenamefont {Saraiva},
  \citenamefont {Tanttu}, \citenamefont {Huang}, \citenamefont
  {Camirand~Lemyre}, \citenamefont {Chan}, \citenamefont {Tan}, \citenamefont
  {Hudson}, \citenamefont {Itoh}, \citenamefont {Morello}, \citenamefont
  {Pioro-Ladri\'{e}re}, \citenamefont {Laucht},\ and\ \citenamefont
  {Dzurak}}]{Yang:2020p350}%
  \BibitemOpen
  \bibfield  {author} {\bibinfo {author} {\bibfnamefont {C.~H.}\ \bibnamefont
  {Yang}}, \bibinfo {author} {\bibfnamefont {R.~C.}\ \bibnamefont {Leon}},
  \bibinfo {author} {\bibfnamefont {J.~C.}\ \bibnamefont {Hwang}}, \bibinfo
  {author} {\bibfnamefont {A.}~\bibnamefont {Saraiva}}, \bibinfo {author}
  {\bibfnamefont {T.}~\bibnamefont {Tanttu}}, \bibinfo {author} {\bibfnamefont
  {W.}~\bibnamefont {Huang}}, \bibinfo {author} {\bibfnamefont
  {J.}~\bibnamefont {Camirand~Lemyre}}, \bibinfo {author} {\bibfnamefont
  {K.~W.}\ \bibnamefont {Chan}}, \bibinfo {author} {\bibfnamefont {K.~Y.}\
  \bibnamefont {Tan}}, \bibinfo {author} {\bibfnamefont {F.~E.}\ \bibnamefont
  {Hudson}}, \bibinfo {author} {\bibfnamefont {K.~M.}\ \bibnamefont {Itoh}},
  \bibinfo {author} {\bibfnamefont {A.}~\bibnamefont {Morello}}, \bibinfo
  {author} {\bibfnamefont {M.}~\bibnamefont {Pioro-Ladri\'{e}re}}, \bibinfo
  {author} {\bibfnamefont {A.}~\bibnamefont {Laucht}}, \ and\ \bibinfo {author}
  {\bibfnamefont {A.~S.}\ \bibnamefont {Dzurak}},\ }\href {\doibase
  10.1038/s41586-020-2171-6} {\bibfield  {journal} {\bibinfo  {journal}
  {Nature}\ }\textbf {\bibinfo {volume} {580}},\ \bibinfo {pages} {350}
  (\bibinfo {year} {2020})}\BibitemShut {NoStop}%
\bibitem [{\citenamefont {Leon}\ \emph {et~al.}(2020)\citenamefont {Leon},
  \citenamefont {Yang}, \citenamefont {Hwang}, \citenamefont {Lemyre},
  \citenamefont {Tanttu}, \citenamefont {Huang}, \citenamefont {Chan},
  \citenamefont {Tan}, \citenamefont {Hudson}, \citenamefont {Itoh},
  \citenamefont {Morello}, \citenamefont {Laucht}, \citenamefont
  {Pioro-Ladri\'{e}re}, \citenamefont {Saraiva},\ and\ \citenamefont
  {Dzurak}}]{Leon:2020p797}%
  \BibitemOpen
  \bibfield  {author} {\bibinfo {author} {\bibfnamefont {R.~C.}\ \bibnamefont
  {Leon}}, \bibinfo {author} {\bibfnamefont {C.~H.}\ \bibnamefont {Yang}},
  \bibinfo {author} {\bibfnamefont {J.~C.}\ \bibnamefont {Hwang}}, \bibinfo
  {author} {\bibfnamefont {J.~C.}\ \bibnamefont {Lemyre}}, \bibinfo {author}
  {\bibfnamefont {T.}~\bibnamefont {Tanttu}}, \bibinfo {author} {\bibfnamefont
  {W.}~\bibnamefont {Huang}}, \bibinfo {author} {\bibfnamefont {K.~W.}\
  \bibnamefont {Chan}}, \bibinfo {author} {\bibfnamefont {K.~Y.}\ \bibnamefont
  {Tan}}, \bibinfo {author} {\bibfnamefont {F.~E.}\ \bibnamefont {Hudson}},
  \bibinfo {author} {\bibfnamefont {K.~M.}\ \bibnamefont {Itoh}}, \bibinfo
  {author} {\bibfnamefont {A.}~\bibnamefont {Morello}}, \bibinfo {author}
  {\bibfnamefont {A.}~\bibnamefont {Laucht}}, \bibinfo {author} {\bibfnamefont
  {M.}~\bibnamefont {Pioro-Ladri\'{e}re}}, \bibinfo {author} {\bibfnamefont
  {A.}~\bibnamefont {Saraiva}}, \ and\ \bibinfo {author} {\bibfnamefont
  {A.~S.}\ \bibnamefont {Dzurak}},\ }\href {\doibase
  10.1038/s41467-019-14053-w} {\bibfield  {journal} {\bibinfo  {journal} {Nat.
  Commun.}\ }\textbf {\bibinfo {volume} {11}},\ \bibinfo {pages} {797}
  (\bibinfo {year} {2020})}\BibitemShut {NoStop}%
\bibitem [{\citenamefont {Kim}\ \emph {et~al.}(2014)\citenamefont {Kim},
  \citenamefont {Shi}, \citenamefont {Simmons}, \citenamefont {Ward},
  \citenamefont {Prance}, \citenamefont {Koh}, \citenamefont {Gamble},
  \citenamefont {Savage}, \citenamefont {Lagally}, \citenamefont {Friesen},
  \citenamefont {Coppersmith},\ and\ \citenamefont {Eriksson}}]{Kim:2014p70}%
  \BibitemOpen
  \bibfield  {author} {\bibinfo {author} {\bibfnamefont {D.}~\bibnamefont
  {Kim}}, \bibinfo {author} {\bibfnamefont {Z.}~\bibnamefont {Shi}}, \bibinfo
  {author} {\bibfnamefont {C.~B.}\ \bibnamefont {Simmons}}, \bibinfo {author}
  {\bibfnamefont {D.~R.}\ \bibnamefont {Ward}}, \bibinfo {author}
  {\bibfnamefont {J.~R.}\ \bibnamefont {Prance}}, \bibinfo {author}
  {\bibfnamefont {T.~S.}\ \bibnamefont {Koh}}, \bibinfo {author} {\bibfnamefont
  {J.~K.}\ \bibnamefont {Gamble}}, \bibinfo {author} {\bibfnamefont {D.~E.}\
  \bibnamefont {Savage}}, \bibinfo {author} {\bibfnamefont {M.~G.}\
  \bibnamefont {Lagally}}, \bibinfo {author} {\bibfnamefont {M.}~\bibnamefont
  {Friesen}}, \bibinfo {author} {\bibfnamefont {S.~N.}\ \bibnamefont
  {Coppersmith}}, \ and\ \bibinfo {author} {\bibfnamefont {M.~A.}\ \bibnamefont
  {Eriksson}},\ }\href {\doibase 10.1038/nature13407} {\bibfield  {journal}
  {\bibinfo  {journal} {Nature}\ }\textbf {\bibinfo {volume} {511}},\ \bibinfo
  {pages} {70} (\bibinfo {year} {2014})}\BibitemShut {NoStop}%
\bibitem [{\citenamefont {Ono}\ \emph {et~al.}(2002)\citenamefont {Ono},
  \citenamefont {Austing}, \citenamefont {Tokura},\ and\ \citenamefont
  {Tarucha}}]{Ono:2002p1313}%
  \BibitemOpen
  \bibfield  {author} {\bibinfo {author} {\bibfnamefont {K.}~\bibnamefont
  {Ono}}, \bibinfo {author} {\bibfnamefont {D.~G.}\ \bibnamefont {Austing}},
  \bibinfo {author} {\bibfnamefont {Y.}~\bibnamefont {Tokura}}, \ and\ \bibinfo
  {author} {\bibfnamefont {S.}~\bibnamefont {Tarucha}},\ }\href {\doibase
  10.1126/science.1070958} {\bibfield  {journal} {\bibinfo  {journal}
  {Science}\ }\textbf {\bibinfo {volume} {297}},\ \bibinfo {pages} {1313}
  (\bibinfo {year} {2002})}\BibitemShut {NoStop}%
\bibitem [{\citenamefont {Petta}\ \emph {et~al.}(2005)\citenamefont {Petta},
  \citenamefont {Johnson}, \citenamefont {Taylor}, \citenamefont {Laird},
  \citenamefont {Yacoby}, \citenamefont {Lukin}, \citenamefont {Marcus},
  \citenamefont {Hanson},\ and\ \citenamefont {Gossard}}]{Petta:2005p2180}%
  \BibitemOpen
  \bibfield  {author} {\bibinfo {author} {\bibfnamefont {J.~R.}\ \bibnamefont
  {Petta}}, \bibinfo {author} {\bibfnamefont {A.~C.}\ \bibnamefont {Johnson}},
  \bibinfo {author} {\bibfnamefont {J.~M.}\ \bibnamefont {Taylor}}, \bibinfo
  {author} {\bibfnamefont {E.~A.}\ \bibnamefont {Laird}}, \bibinfo {author}
  {\bibfnamefont {A.}~\bibnamefont {Yacoby}}, \bibinfo {author} {\bibfnamefont
  {M.~D.}\ \bibnamefont {Lukin}}, \bibinfo {author} {\bibfnamefont {C.~M.}\
  \bibnamefont {Marcus}}, \bibinfo {author} {\bibfnamefont {M.~P.}\
  \bibnamefont {Hanson}}, \ and\ \bibinfo {author} {\bibfnamefont {A.~C.}\
  \bibnamefont {Gossard}},\ }\href {\doibase 10.1126/science.1116955}
  {\bibfield  {journal} {\bibinfo  {journal} {Science}\ }\textbf {\bibinfo
  {volume} {309}},\ \bibinfo {pages} {2180} (\bibinfo {year}
  {2005})}\BibitemShut {NoStop}%
\bibitem [{\citenamefont {Shulman}\ \emph {et~al.}(2012)\citenamefont
  {Shulman}, \citenamefont {Dial}, \citenamefont {Harvey}, \citenamefont
  {Bluhm}, \citenamefont {Umansky},\ and\ \citenamefont
  {Yacoby}}]{Shulman:2012p202}%
  \BibitemOpen
  \bibfield  {author} {\bibinfo {author} {\bibfnamefont {M.~D.}\ \bibnamefont
  {Shulman}}, \bibinfo {author} {\bibfnamefont {O.~E.}\ \bibnamefont {Dial}},
  \bibinfo {author} {\bibfnamefont {S.~P.}\ \bibnamefont {Harvey}}, \bibinfo
  {author} {\bibfnamefont {H.}~\bibnamefont {Bluhm}}, \bibinfo {author}
  {\bibfnamefont {V.}~\bibnamefont {Umansky}}, \ and\ \bibinfo {author}
  {\bibfnamefont {A.}~\bibnamefont {Yacoby}},\ }\href@noop {} {\bibfield
  {journal} {\bibinfo  {journal} {Science}\ }\textbf {\bibinfo {volume}
  {336}},\ \bibinfo {pages} {202} (\bibinfo {year} {2012})}\BibitemShut
  {NoStop}%
\bibitem [{\citenamefont {Reed}\ \emph {et~al.}(2016)\citenamefont {Reed},
  \citenamefont {Maune}, \citenamefont {Andrews}, \citenamefont {Borselli},
  \citenamefont {Eng}, \citenamefont {Jura}, \citenamefont {Kiselev},
  \citenamefont {Ladd}, \citenamefont {Merkel}, \citenamefont {Milosavljevic},
  \citenamefont {Pritchett}, \citenamefont {Rakher}, \citenamefont {Ross},
  \citenamefont {Schmitz}, \citenamefont {Smith}, \citenamefont {Wright},
  \citenamefont {Gyure},\ and\ \citenamefont {Hunter}}]{Reed:2016p110402}%
  \BibitemOpen
  \bibfield  {author} {\bibinfo {author} {\bibfnamefont {M.~D.}\ \bibnamefont
  {Reed}}, \bibinfo {author} {\bibfnamefont {B.~M.}\ \bibnamefont {Maune}},
  \bibinfo {author} {\bibfnamefont {R.~W.}\ \bibnamefont {Andrews}}, \bibinfo
  {author} {\bibfnamefont {M.~G.}\ \bibnamefont {Borselli}}, \bibinfo {author}
  {\bibfnamefont {K.}~\bibnamefont {Eng}}, \bibinfo {author} {\bibfnamefont
  {M.~P.}\ \bibnamefont {Jura}}, \bibinfo {author} {\bibfnamefont {A.~A.}\
  \bibnamefont {Kiselev}}, \bibinfo {author} {\bibfnamefont {T.~D.}\
  \bibnamefont {Ladd}}, \bibinfo {author} {\bibfnamefont {S.~T.}\ \bibnamefont
  {Merkel}}, \bibinfo {author} {\bibfnamefont {I.}~\bibnamefont
  {Milosavljevic}}, \bibinfo {author} {\bibfnamefont {E.~J.}\ \bibnamefont
  {Pritchett}}, \bibinfo {author} {\bibfnamefont {M.~T.}\ \bibnamefont
  {Rakher}}, \bibinfo {author} {\bibfnamefont {R.~S.}\ \bibnamefont {Ross}},
  \bibinfo {author} {\bibfnamefont {A.~E.}\ \bibnamefont {Schmitz}}, \bibinfo
  {author} {\bibfnamefont {A.}~\bibnamefont {Smith}}, \bibinfo {author}
  {\bibfnamefont {J.~A.}\ \bibnamefont {Wright}}, \bibinfo {author}
  {\bibfnamefont {M.~F.}\ \bibnamefont {Gyure}}, \ and\ \bibinfo {author}
  {\bibfnamefont {A.~T.}\ \bibnamefont {Hunter}},\ }\href {\doibase
  10.1103/PhysRevLett.116.110402} {\bibfield  {journal} {\bibinfo  {journal}
  {Phys. Rev. Lett.}\ }\textbf {\bibinfo {volume} {116}},\ \bibinfo {pages}
  {110402} (\bibinfo {year} {2016})}\BibitemShut {NoStop}%
\bibitem [{\citenamefont {Gaudreau}\ \emph {et~al.}(2012)\citenamefont
  {Gaudreau}, \citenamefont {Granger}, \citenamefont {Kam}, \citenamefont
  {Aers}, \citenamefont {Studenikin}, \citenamefont {Zawadzki}, \citenamefont
  {Pioro-Ladri\`{e}re}, \citenamefont {Wasilewski},\ and\ \citenamefont
  {Sachrajda}}]{Gaudreau:2011p54}%
  \BibitemOpen
  \bibfield  {author} {\bibinfo {author} {\bibfnamefont {L.}~\bibnamefont
  {Gaudreau}}, \bibinfo {author} {\bibfnamefont {G.}~\bibnamefont {Granger}},
  \bibinfo {author} {\bibfnamefont {A.}~\bibnamefont {Kam}}, \bibinfo {author}
  {\bibfnamefont {G.~C.}\ \bibnamefont {Aers}}, \bibinfo {author}
  {\bibfnamefont {S.~A.}\ \bibnamefont {Studenikin}}, \bibinfo {author}
  {\bibfnamefont {P.}~\bibnamefont {Zawadzki}}, \bibinfo {author}
  {\bibfnamefont {M.}~\bibnamefont {Pioro-Ladri\`{e}re}}, \bibinfo {author}
  {\bibfnamefont {Z.~R.}\ \bibnamefont {Wasilewski}}, \ and\ \bibinfo {author}
  {\bibfnamefont {A.~S.}\ \bibnamefont {Sachrajda}},\ }\href {\doibase
  10.1038/nphys2149} {\bibfield  {journal} {\bibinfo  {journal} {Nat. Phys.}\
  }\textbf {\bibinfo {volume} {8}},\ \bibinfo {pages} {54} (\bibinfo {year}
  {2012})}\BibitemShut {NoStop}%
\bibitem [{\citenamefont {Medford}\ \emph {et~al.}(2013)\citenamefont
  {Medford}, \citenamefont {Beil}, \citenamefont {Taylor}, \citenamefont
  {Rashba}, \citenamefont {Lu}, \citenamefont {Gossard},\ and\ \citenamefont
  {Marcus}}]{Medford:2013p050501}%
  \BibitemOpen
  \bibfield  {author} {\bibinfo {author} {\bibfnamefont {J.}~\bibnamefont
  {Medford}}, \bibinfo {author} {\bibfnamefont {J.}~\bibnamefont {Beil}},
  \bibinfo {author} {\bibfnamefont {J.~M.}\ \bibnamefont {Taylor}}, \bibinfo
  {author} {\bibfnamefont {E.~I.}\ \bibnamefont {Rashba}}, \bibinfo {author}
  {\bibfnamefont {H.}~\bibnamefont {Lu}}, \bibinfo {author} {\bibfnamefont
  {A.~C.}\ \bibnamefont {Gossard}}, \ and\ \bibinfo {author} {\bibfnamefont
  {C.~M.}\ \bibnamefont {Marcus}},\ }\href {\doibase
  10.1103/PhysRevLett.111.050501} {\bibfield  {journal} {\bibinfo  {journal}
  {Phys. Rev. Lett.}\ }\textbf {\bibinfo {volume} {111}},\ \bibinfo {pages}
  {050501} (\bibinfo {year} {2013})}\BibitemShut {NoStop}%
\bibitem [{\citenamefont {Shi}\ \emph {et~al.}(2012)\citenamefont {Shi},
  \citenamefont {Simmons}, \citenamefont {Prance}, \citenamefont {Gamble},
  \citenamefont {Koh}, \citenamefont {Shim}, \citenamefont {Hu}, \citenamefont
  {Savage}, \citenamefont {Lagally}, \citenamefont {Eriksson}, \citenamefont
  {Friesen},\ and\ \citenamefont {Coppersmith}}]{Shi:2012p140503}%
  \BibitemOpen
  \bibfield  {author} {\bibinfo {author} {\bibfnamefont {Z.}~\bibnamefont
  {Shi}}, \bibinfo {author} {\bibfnamefont {C.~B.}\ \bibnamefont {Simmons}},
  \bibinfo {author} {\bibfnamefont {J.~R.}\ \bibnamefont {Prance}}, \bibinfo
  {author} {\bibfnamefont {J.~K.}\ \bibnamefont {Gamble}}, \bibinfo {author}
  {\bibfnamefont {T.~S.}\ \bibnamefont {Koh}}, \bibinfo {author} {\bibfnamefont
  {Y.-P.}\ \bibnamefont {Shim}}, \bibinfo {author} {\bibfnamefont
  {X.}~\bibnamefont {Hu}}, \bibinfo {author} {\bibfnamefont {D.~E.}\
  \bibnamefont {Savage}}, \bibinfo {author} {\bibfnamefont {M.~G.}\
  \bibnamefont {Lagally}}, \bibinfo {author} {\bibfnamefont {M.~A.}\
  \bibnamefont {Eriksson}}, \bibinfo {author} {\bibfnamefont {M.}~\bibnamefont
  {Friesen}}, \ and\ \bibinfo {author} {\bibfnamefont {S.~N.}\ \bibnamefont
  {Coppersmith}},\ }\href {\doibase 10.1103/PhysRevLett.108.140503} {\bibfield
  {journal} {\bibinfo  {journal} {Phys. Rev. Lett.}\ }\textbf {\bibinfo
  {volume} {108}},\ \bibinfo {pages} {140503} (\bibinfo {year}
  {2012})}\BibitemShut {NoStop}%
\bibitem [{\citenamefont {Li}\ \emph {et~al.}(2018)\citenamefont {Li},
  \citenamefont {Petit}, \citenamefont {Franke}, \citenamefont {Dehollain},
  \citenamefont {Helsen}, \citenamefont {Steudtner}, \citenamefont {Thomas},
  \citenamefont {Yoscovits}, \citenamefont {Singh}, \citenamefont {Wehner},
  \citenamefont {Vandersypen}, \citenamefont {Clarke},\ and\ \citenamefont
  {Veldhorst}}]{Li:2018p7}%
  \BibitemOpen
  \bibfield  {author} {\bibinfo {author} {\bibfnamefont {R.}~\bibnamefont
  {Li}}, \bibinfo {author} {\bibfnamefont {L.}~\bibnamefont {Petit}}, \bibinfo
  {author} {\bibfnamefont {D.~P.}\ \bibnamefont {Franke}}, \bibinfo {author}
  {\bibfnamefont {J.~P.}\ \bibnamefont {Dehollain}}, \bibinfo {author}
  {\bibfnamefont {J.}~\bibnamefont {Helsen}}, \bibinfo {author} {\bibfnamefont
  {M.}~\bibnamefont {Steudtner}}, \bibinfo {author} {\bibfnamefont {N.~K.}\
  \bibnamefont {Thomas}}, \bibinfo {author} {\bibfnamefont {Z.~R.}\
  \bibnamefont {Yoscovits}}, \bibinfo {author} {\bibfnamefont {K.~J.}\
  \bibnamefont {Singh}}, \bibinfo {author} {\bibfnamefont {S.}~\bibnamefont
  {Wehner}}, \bibinfo {author} {\bibfnamefont {L.~M.~K.}\ \bibnamefont
  {Vandersypen}}, \bibinfo {author} {\bibfnamefont {J.~S.}\ \bibnamefont
  {Clarke}}, \ and\ \bibinfo {author} {\bibfnamefont {M.}~\bibnamefont
  {Veldhorst}},\ }\href@noop {} {\bibfield  {journal} {\bibinfo  {journal}
  {Science Advances}\ }\textbf {\bibinfo {volume} {4}},\ \bibinfo {pages} {7}
  (\bibinfo {year} {2018})}\BibitemShut {NoStop}%
\bibitem [{\citenamefont {Vandersypen}\ \emph {et~al.}(2017)\citenamefont
  {Vandersypen}, \citenamefont {Bluhm}, \citenamefont {Clarke}, \citenamefont
  {Dzurak}, \citenamefont {Ishihara}, \citenamefont {Morello}, \citenamefont
  {Reilly}, \citenamefont {Schreiber},\ and\ \citenamefont
  {Veldhorst}}]{Vandersypen:2017p34}%
  \BibitemOpen
  \bibfield  {author} {\bibinfo {author} {\bibfnamefont {L.~M.~K.}\
  \bibnamefont {Vandersypen}}, \bibinfo {author} {\bibfnamefont
  {H.}~\bibnamefont {Bluhm}}, \bibinfo {author} {\bibfnamefont {J.~S.}\
  \bibnamefont {Clarke}}, \bibinfo {author} {\bibfnamefont {A.~S.}\
  \bibnamefont {Dzurak}}, \bibinfo {author} {\bibfnamefont {R.}~\bibnamefont
  {Ishihara}}, \bibinfo {author} {\bibfnamefont {A.}~\bibnamefont {Morello}},
  \bibinfo {author} {\bibfnamefont {D.~J.}\ \bibnamefont {Reilly}}, \bibinfo
  {author} {\bibfnamefont {L.~R.}\ \bibnamefont {Schreiber}}, \ and\ \bibinfo
  {author} {\bibfnamefont {M.}~\bibnamefont {Veldhorst}},\ }\href {\doibase
  10.1038/s41534-017-0038-y} {\bibfield  {journal} {\bibinfo  {journal} {npj
  Quantum Inf.}\ }\textbf {\bibinfo {volume} {3}},\ \bibinfo {pages} {34}
  (\bibinfo {year} {2017})}\BibitemShut {NoStop}%
\bibitem [{\citenamefont {Petit}\ \emph {et~al.}(2020)\citenamefont {Petit},
  \citenamefont {Eenink}, \citenamefont {Russ}, \citenamefont {Lawrie},
  \citenamefont {Hendrickx}, \citenamefont {Philips}, \citenamefont {Clarke},
  \citenamefont {Vandersypen},\ and\ \citenamefont
  {Veldhorst}}]{Petit:2020p355}%
  \BibitemOpen
  \bibfield  {author} {\bibinfo {author} {\bibfnamefont {L.}~\bibnamefont
  {Petit}}, \bibinfo {author} {\bibfnamefont {H.~G.~J.}\ \bibnamefont
  {Eenink}}, \bibinfo {author} {\bibfnamefont {M.}~\bibnamefont {Russ}},
  \bibinfo {author} {\bibfnamefont {W.~I.~L.}\ \bibnamefont {Lawrie}}, \bibinfo
  {author} {\bibfnamefont {N.~W.}\ \bibnamefont {Hendrickx}}, \bibinfo {author}
  {\bibfnamefont {S.~G.~J.}\ \bibnamefont {Philips}}, \bibinfo {author}
  {\bibfnamefont {J.~S.}\ \bibnamefont {Clarke}}, \bibinfo {author}
  {\bibfnamefont {L.~M.~K.}\ \bibnamefont {Vandersypen}}, \ and\ \bibinfo
  {author} {\bibfnamefont {M.}~\bibnamefont {Veldhorst}},\ }\href {\doibase
  10.1038/s41586-020-2170-7} {\bibfield  {journal} {\bibinfo  {journal}
  {Nature}\ }\textbf {\bibinfo {volume} {580}},\ \bibinfo {pages} {355}
  (\bibinfo {year} {2020})}\BibitemShut {NoStop}%
\bibitem [{\citenamefont {Bryant}(1987)}]{Bryant:1987p1140}%
  \BibitemOpen
  \bibfield  {author} {\bibinfo {author} {\bibfnamefont {G.~W.}\ \bibnamefont
  {Bryant}},\ }\href {\doibase 10.1103/PhysRevLett.59.1140} {\bibfield
  {journal} {\bibinfo  {journal} {Phys. Rev. Lett.}\ }\textbf {\bibinfo
  {volume} {59}},\ \bibinfo {pages} {1140} (\bibinfo {year}
  {1987})}\BibitemShut {NoStop}%
\bibitem [{\citenamefont {Yannouleas}\ and\ \citenamefont
  {Landman}(1999)}]{Yannouleas:1999p5325}%
  \BibitemOpen
  \bibfield  {author} {\bibinfo {author} {\bibfnamefont {C.}~\bibnamefont
  {Yannouleas}}\ and\ \bibinfo {author} {\bibfnamefont {U.}~\bibnamefont
  {Landman}},\ }\href {\doibase 10.1103/PhysRevLett.82.5325} {\bibfield
  {journal} {\bibinfo  {journal} {Phys. Rev. Lett.}\ }\textbf {\bibinfo
  {volume} {82}},\ \bibinfo {pages} {5325} (\bibinfo {year}
  {1999})}\BibitemShut {NoStop}%
\bibitem [{\citenamefont {Reimann}\ \emph {et~al.}(2000)\citenamefont
  {Reimann}, \citenamefont {Koskinen},\ and\ \citenamefont
  {Manninen}}]{Reimann:2000p8108}%
  \BibitemOpen
  \bibfield  {author} {\bibinfo {author} {\bibfnamefont {S.~M.}\ \bibnamefont
  {Reimann}}, \bibinfo {author} {\bibfnamefont {M.}~\bibnamefont {Koskinen}}, \
  and\ \bibinfo {author} {\bibfnamefont {M.}~\bibnamefont {Manninen}},\ }\href
  {\doibase 10.1103/PhysRevB.62.8108} {\bibfield  {journal} {\bibinfo
  {journal} {Phys. Rev. B}\ }\textbf {\bibinfo {volume} {62}},\ \bibinfo
  {pages} {8108} (\bibinfo {year} {2000})}\BibitemShut {NoStop}%
\bibitem [{\citenamefont {Reusch}\ \emph {et~al.}(2001)\citenamefont {Reusch},
  \citenamefont {H{\"a}usler},\ and\ \citenamefont
  {Grabert}}]{Reusch:2001p113313}%
  \BibitemOpen
  \bibfield  {author} {\bibinfo {author} {\bibfnamefont {B.}~\bibnamefont
  {Reusch}}, \bibinfo {author} {\bibfnamefont {W.}~\bibnamefont {H{\"a}usler}},
  \ and\ \bibinfo {author} {\bibfnamefont {H.}~\bibnamefont {Grabert}},\ }\href
  {\doibase 10.1103/PhysRevB.63.113313} {\bibfield  {journal} {\bibinfo
  {journal} {Phys. Rev. B}\ }\textbf {\bibinfo {volume} {63}},\ \bibinfo
  {pages} {113313} (\bibinfo {year} {2001})}\BibitemShut {NoStop}%
\bibitem [{\citenamefont {Rontani}\ \emph {et~al.}(2006)\citenamefont
  {Rontani}, \citenamefont {Cavazzoni}, \citenamefont {Bellucci},\ and\
  \citenamefont {Goldoni}}]{Rontani:2006p124102}%
  \BibitemOpen
  \bibfield  {author} {\bibinfo {author} {\bibfnamefont {M.}~\bibnamefont
  {Rontani}}, \bibinfo {author} {\bibfnamefont {C.}~\bibnamefont {Cavazzoni}},
  \bibinfo {author} {\bibfnamefont {D.}~\bibnamefont {Bellucci}}, \ and\
  \bibinfo {author} {\bibfnamefont {G.}~\bibnamefont {Goldoni}},\ }\href
  {\doibase 10.1063/1.2179418} {\bibfield  {journal} {\bibinfo  {journal} {The
  Journal of Chemical Physics}\ }\textbf {\bibinfo {volume} {124}},\ \bibinfo
  {pages} {124102} (\bibinfo {year} {2006})}\BibitemShut {NoStop}%
\bibitem [{\citenamefont {Ghosal}\ \emph {et~al.}(2007)\citenamefont {Ghosal},
  \citenamefont {G\"u\ifmmode~\mbox{\c{c}}\else \c{c}\fi{}l\"u}, \citenamefont
  {Umrigar}, \citenamefont {Ullmo},\ and\ \citenamefont
  {Baranger}}]{Ghosal:2007p085341}%
  \BibitemOpen
  \bibfield  {author} {\bibinfo {author} {\bibfnamefont {A.}~\bibnamefont
  {Ghosal}}, \bibinfo {author} {\bibfnamefont {A.~D.}\ \bibnamefont
  {G\"u\ifmmode~\mbox{\c{c}}\else \c{c}\fi{}l\"u}}, \bibinfo {author}
  {\bibfnamefont {C.~J.}\ \bibnamefont {Umrigar}}, \bibinfo {author}
  {\bibfnamefont {D.}~\bibnamefont {Ullmo}}, \ and\ \bibinfo {author}
  {\bibfnamefont {H.~U.}\ \bibnamefont {Baranger}},\ }\href {\doibase
  10.1103/PhysRevB.76.085341} {\bibfield  {journal} {\bibinfo  {journal} {Phys.
  Rev. B}\ }\textbf {\bibinfo {volume} {76}},\ \bibinfo {pages} {085341}
  (\bibinfo {year} {2007})}\BibitemShut {NoStop}%
\bibitem [{\citenamefont {Shapir}\ \emph {et~al.}(2019)\citenamefont {Shapir},
  \citenamefont {Hamo}, \citenamefont {Pecker}, \citenamefont {Moca},
  \citenamefont {Legeza}, \citenamefont {Zarand},\ and\ \citenamefont
  {Ilani}}]{Shapir:2019p870}%
  \BibitemOpen
  \bibfield  {author} {\bibinfo {author} {\bibfnamefont {I.}~\bibnamefont
  {Shapir}}, \bibinfo {author} {\bibfnamefont {A.}~\bibnamefont {Hamo}},
  \bibinfo {author} {\bibfnamefont {S.}~\bibnamefont {Pecker}}, \bibinfo
  {author} {\bibfnamefont {C.~P.}\ \bibnamefont {Moca}}, \bibinfo {author}
  {\bibfnamefont {{\"O}.}~\bibnamefont {Legeza}}, \bibinfo {author}
  {\bibfnamefont {G.}~\bibnamefont {Zarand}}, \ and\ \bibinfo {author}
  {\bibfnamefont {S.}~\bibnamefont {Ilani}},\ }\href {\doibase
  10.1126/science.aat0905} {\bibfield  {journal} {\bibinfo  {journal}
  {Science}\ }\textbf {\bibinfo {volume} {364}},\ \bibinfo {pages} {870}
  (\bibinfo {year} {2019})}\BibitemShut {NoStop}%
\bibitem [{\citenamefont {Mintairov}\ \emph {et~al.}(2018)\citenamefont
  {Mintairov}, \citenamefont {Kapaldo}, \citenamefont {Merz}, \citenamefont
  {Rouvimov}, \citenamefont {Lebedev}, \citenamefont {Kalyuzhnyy},
  \citenamefont {Mintairov}, \citenamefont {Belyaev}, \citenamefont {Rakhlin},
  \citenamefont {Toropov}, \citenamefont {Brunkov}, \citenamefont {Vlasov},
  \citenamefont {Zadiranov}, \citenamefont {Blundell}, \citenamefont
  {Mozharov}, \citenamefont {Mukhin}, \citenamefont {Yakimov}, \citenamefont
  {Oktyabrsky}, \citenamefont {Shelaev},\ and\ \citenamefont
  {Bykov}}]{Mintairov:2018p195443}%
  \BibitemOpen
  \bibfield  {author} {\bibinfo {author} {\bibfnamefont {A.~M.}\ \bibnamefont
  {Mintairov}}, \bibinfo {author} {\bibfnamefont {J.}~\bibnamefont {Kapaldo}},
  \bibinfo {author} {\bibfnamefont {J.~L.}\ \bibnamefont {Merz}}, \bibinfo
  {author} {\bibfnamefont {S.}~\bibnamefont {Rouvimov}}, \bibinfo {author}
  {\bibfnamefont {D.~V.}\ \bibnamefont {Lebedev}}, \bibinfo {author}
  {\bibfnamefont {N.~A.}\ \bibnamefont {Kalyuzhnyy}}, \bibinfo {author}
  {\bibfnamefont {S.~A.}\ \bibnamefont {Mintairov}}, \bibinfo {author}
  {\bibfnamefont {K.~G.}\ \bibnamefont {Belyaev}}, \bibinfo {author}
  {\bibfnamefont {M.~V.}\ \bibnamefont {Rakhlin}}, \bibinfo {author}
  {\bibfnamefont {A.~A.}\ \bibnamefont {Toropov}}, \bibinfo {author}
  {\bibfnamefont {P.~N.}\ \bibnamefont {Brunkov}}, \bibinfo {author}
  {\bibfnamefont {A.~S.}\ \bibnamefont {Vlasov}}, \bibinfo {author}
  {\bibfnamefont {Y.}~\bibnamefont {Zadiranov}}, \bibinfo {author}
  {\bibfnamefont {S.~A.}\ \bibnamefont {Blundell}}, \bibinfo {author}
  {\bibfnamefont {A.~M.}\ \bibnamefont {Mozharov}}, \bibinfo {author}
  {\bibfnamefont {I.}~\bibnamefont {Mukhin}}, \bibinfo {author} {\bibfnamefont
  {M.}~\bibnamefont {Yakimov}}, \bibinfo {author} {\bibfnamefont
  {S.}~\bibnamefont {Oktyabrsky}}, \bibinfo {author} {\bibfnamefont {A.~V.}\
  \bibnamefont {Shelaev}}, \ and\ \bibinfo {author} {\bibfnamefont {V.~A.}\
  \bibnamefont {Bykov}},\ }\href {\doibase 10.1103/PhysRevB.97.195443}
  {\bibfield  {journal} {\bibinfo  {journal} {Phys. Rev. B}\ }\textbf {\bibinfo
  {volume} {97}},\ \bibinfo {pages} {195443} (\bibinfo {year}
  {2018})}\BibitemShut {NoStop}%
\bibitem [{\citenamefont {Kalliakos}\ \emph {et~al.}(2008)\citenamefont
  {Kalliakos}, \citenamefont {Rontani}, \citenamefont {Pellegrini},
  \citenamefont {Garc\'ia}, \citenamefont {Pinczuk}, \citenamefont {Goldoni},
  \citenamefont {ad~Loren N.~Pfeiffer},\ and\ \citenamefont
  {West}}]{Kalliakos:2008p467}%
  \BibitemOpen
  \bibfield  {author} {\bibinfo {author} {\bibfnamefont {S.}~\bibnamefont
  {Kalliakos}}, \bibinfo {author} {\bibfnamefont {M.}~\bibnamefont {Rontani}},
  \bibinfo {author} {\bibfnamefont {V.}~\bibnamefont {Pellegrini}}, \bibinfo
  {author} {\bibfnamefont {C.~P.}\ \bibnamefont {Garc\'ia}}, \bibinfo {author}
  {\bibfnamefont {A.}~\bibnamefont {Pinczuk}}, \bibinfo {author} {\bibfnamefont
  {G.}~\bibnamefont {Goldoni}}, \bibinfo {author} {\bibfnamefont {E.~M.}\
  \bibnamefont {ad~Loren N.~Pfeiffer}}, \ and\ \bibinfo {author} {\bibfnamefont
  {K.~W.}\ \bibnamefont {West}},\ }\href {\doibase 10.1038/nphys944} {\bibfield
   {journal} {\bibinfo  {journal} {Nat. Phys.}\ }\textbf {\bibinfo {volume}
  {4}},\ \bibinfo {pages} {467} (\bibinfo {year} {2008})}\BibitemShut {NoStop}%
\bibitem [{\citenamefont {Singha}\ \emph {et~al.}(2010)\citenamefont {Singha},
  \citenamefont {Pellegrini}, \citenamefont {Pinczuk}, \citenamefont
  {Pfeiffer}, \citenamefont {West},\ and\ \citenamefont
  {Rontani}}]{Singha:2010p246802}%
  \BibitemOpen
  \bibfield  {author} {\bibinfo {author} {\bibfnamefont {A.}~\bibnamefont
  {Singha}}, \bibinfo {author} {\bibfnamefont {V.}~\bibnamefont {Pellegrini}},
  \bibinfo {author} {\bibfnamefont {A.}~\bibnamefont {Pinczuk}}, \bibinfo
  {author} {\bibfnamefont {L.~N.}\ \bibnamefont {Pfeiffer}}, \bibinfo {author}
  {\bibfnamefont {K.~W.}\ \bibnamefont {West}}, \ and\ \bibinfo {author}
  {\bibfnamefont {M.}~\bibnamefont {Rontani}},\ }\href {\doibase
  10.1103/PhysRevLett.104.246802} {\bibfield  {journal} {\bibinfo  {journal}
  {Phys. Rev. Lett.}\ }\textbf {\bibinfo {volume} {104}},\ \bibinfo {pages}
  {246802} (\bibinfo {year} {2010})}\BibitemShut {NoStop}%
\bibitem [{\citenamefont {Ellenberger}\ \emph {et~al.}(2006)\citenamefont
  {Ellenberger}, \citenamefont {Ihn}, \citenamefont {Yannouleas}, \citenamefont
  {Landman}, \citenamefont {Ensslin}, \citenamefont {Driscoll},\ and\
  \citenamefont {Gossard}}]{Ellenberger:2006p126806}%
  \BibitemOpen
  \bibfield  {author} {\bibinfo {author} {\bibfnamefont {C.}~\bibnamefont
  {Ellenberger}}, \bibinfo {author} {\bibfnamefont {T.}~\bibnamefont {Ihn}},
  \bibinfo {author} {\bibfnamefont {C.}~\bibnamefont {Yannouleas}}, \bibinfo
  {author} {\bibfnamefont {U.}~\bibnamefont {Landman}}, \bibinfo {author}
  {\bibfnamefont {K.}~\bibnamefont {Ensslin}}, \bibinfo {author} {\bibfnamefont
  {D.}~\bibnamefont {Driscoll}}, \ and\ \bibinfo {author} {\bibfnamefont
  {A.~C.}\ \bibnamefont {Gossard}},\ }\href {\doibase
  10.1103/PhysRevLett.96.126806} {\bibfield  {journal} {\bibinfo  {journal}
  {Phys. Rev. Lett.}\ }\textbf {\bibinfo {volume} {96}},\ \bibinfo {pages}
  {126806} (\bibinfo {year} {2006})}\BibitemShut {NoStop}%
\bibitem [{\citenamefont {Kristinsd\'ottir}\ \emph {et~al.}(2011)\citenamefont
  {Kristinsd\'ottir}, \citenamefont {Cremon}, \citenamefont {Nilsson},
  \citenamefont {Xu}, \citenamefont {Samuelson}, \citenamefont {Linke},
  \citenamefont {Wacker},\ and\ \citenamefont
  {Reimann}}]{Kristinsdottir:2011p041101}%
  \BibitemOpen
  \bibfield  {author} {\bibinfo {author} {\bibfnamefont {L.~H.}\ \bibnamefont
  {Kristinsd\'ottir}}, \bibinfo {author} {\bibfnamefont {J.~C.}\ \bibnamefont
  {Cremon}}, \bibinfo {author} {\bibfnamefont {H.~A.}\ \bibnamefont {Nilsson}},
  \bibinfo {author} {\bibfnamefont {H.~Q.}\ \bibnamefont {Xu}}, \bibinfo
  {author} {\bibfnamefont {L.}~\bibnamefont {Samuelson}}, \bibinfo {author}
  {\bibfnamefont {H.}~\bibnamefont {Linke}}, \bibinfo {author} {\bibfnamefont
  {A.}~\bibnamefont {Wacker}}, \ and\ \bibinfo {author} {\bibfnamefont {S.~M.}\
  \bibnamefont {Reimann}},\ }\href {\doibase 10.1103/PhysRevB.83.041101}
  {\bibfield  {journal} {\bibinfo  {journal} {Phys. Rev. B}\ }\textbf {\bibinfo
  {volume} {83}},\ \bibinfo {pages} {041101} (\bibinfo {year}
  {2011})}\BibitemShut {NoStop}%
\bibitem [{\citenamefont {Pecker}\ \emph {et~al.}(2013)\citenamefont {Pecker},
  \citenamefont {Kuemmeth}, \citenamefont {Secchi}, \citenamefont {Rontani},
  \citenamefont {Ralph}, \citenamefont {McEuen},\ and\ \citenamefont
  {Ilani}}]{Pecker:2013p576}%
  \BibitemOpen
  \bibfield  {author} {\bibinfo {author} {\bibfnamefont {S.}~\bibnamefont
  {Pecker}}, \bibinfo {author} {\bibfnamefont {F.}~\bibnamefont {Kuemmeth}},
  \bibinfo {author} {\bibfnamefont {A.}~\bibnamefont {Secchi}}, \bibinfo
  {author} {\bibfnamefont {M.}~\bibnamefont {Rontani}}, \bibinfo {author}
  {\bibfnamefont {D.~C.}\ \bibnamefont {Ralph}}, \bibinfo {author}
  {\bibfnamefont {P.~L.}\ \bibnamefont {McEuen}}, \ and\ \bibinfo {author}
  {\bibfnamefont {S.}~\bibnamefont {Ilani}},\ }\href {\doibase
  10.1038/nphys2692} {\bibfield  {journal} {\bibinfo  {journal} {Nat. Phys.}\
  }\textbf {\bibinfo {volume} {9}},\ \bibinfo {pages} {576} (\bibinfo {year}
  {2013})}\BibitemShut {NoStop}%
\bibitem [{\citenamefont {{Zajac}}\ \emph {et~al.}(2015)\citenamefont
  {{Zajac}}, \citenamefont {{Hazard}}, \citenamefont {{Mi}}, \citenamefont
  {{Wang}},\ and\ \citenamefont {{Petta}}}]{Zajac:2015p223507}%
  \BibitemOpen
  \bibfield  {author} {\bibinfo {author} {\bibfnamefont {D.~M.}\ \bibnamefont
  {{Zajac}}}, \bibinfo {author} {\bibfnamefont {T.~M.}\ \bibnamefont
  {{Hazard}}}, \bibinfo {author} {\bibfnamefont {X.}~\bibnamefont {{Mi}}},
  \bibinfo {author} {\bibfnamefont {K.}~\bibnamefont {{Wang}}}, \ and\ \bibinfo
  {author} {\bibfnamefont {J.~R.}\ \bibnamefont {{Petta}}},\ }\href {\doibase
  10.1063/1.4922249} {\bibfield  {journal} {\bibinfo  {journal} {Appl. Phys.
  Lett.}\ }\textbf {\bibinfo {volume} {106}},\ \bibinfo {eid} {223507}
  (\bibinfo {year} {2015})}\BibitemShut {NoStop}%
\bibitem [{\citenamefont {Dodson}\ \emph {et~al.}(2020)\citenamefont {Dodson},
  \citenamefont {Holman}, \citenamefont {Thorgrimsson}, \citenamefont {Neyens},
  \citenamefont {MacQuarrie}, \citenamefont {McJunkin}, \citenamefont {Foote},
  \citenamefont {Edge}, \citenamefont {Coppersmith},\ and\ \citenamefont
  {Eriksson}}]{Dodson:2020preprint}%
  \BibitemOpen
  \bibfield  {author} {\bibinfo {author} {\bibfnamefont {J.~P.}\ \bibnamefont
  {Dodson}}, \bibinfo {author} {\bibfnamefont {N.}~\bibnamefont {Holman}},
  \bibinfo {author} {\bibfnamefont {B.}~\bibnamefont {Thorgrimsson}}, \bibinfo
  {author} {\bibfnamefont {S.~F.}\ \bibnamefont {Neyens}}, \bibinfo {author}
  {\bibfnamefont {E.~R.}\ \bibnamefont {MacQuarrie}}, \bibinfo {author}
  {\bibfnamefont {T.}~\bibnamefont {McJunkin}}, \bibinfo {author}
  {\bibfnamefont {R.~H.}\ \bibnamefont {Foote}}, \bibinfo {author}
  {\bibfnamefont {L.~F.}\ \bibnamefont {Edge}}, \bibinfo {author}
  {\bibfnamefont {S.~N.}\ \bibnamefont {Coppersmith}}, \ and\ \bibinfo {author}
  {\bibfnamefont {M.~A.}\ \bibnamefont {Eriksson}},\ }\href
  {http://arxiv.org/abs/2004.05683} {\  (\bibinfo {year} {2020})},\ \Eprint
  {http://arxiv.org/abs/2004.05683} {arXiv:2004.05683} \BibitemShut {NoStop}%
\bibitem [{\citenamefont {Tracy}\ \emph {et~al.}(2016)\citenamefont {Tracy},
  \citenamefont {Luhman}, \citenamefont {Carr}, \citenamefont {Bishop},
  \citenamefont {Eyck}, \citenamefont {Pluym}, \citenamefont {Wendt},
  \citenamefont {Lilly},\ and\ \citenamefont {Carroll}}]{Tracy:2016p063101}%
  \BibitemOpen
  \bibfield  {author} {\bibinfo {author} {\bibfnamefont {L.~A.}\ \bibnamefont
  {Tracy}}, \bibinfo {author} {\bibfnamefont {D.~R.}\ \bibnamefont {Luhman}},
  \bibinfo {author} {\bibfnamefont {S.~M.}\ \bibnamefont {Carr}}, \bibinfo
  {author} {\bibfnamefont {N.~C.}\ \bibnamefont {Bishop}}, \bibinfo {author}
  {\bibfnamefont {G.~A.~T.}\ \bibnamefont {Eyck}}, \bibinfo {author}
  {\bibfnamefont {T.}~\bibnamefont {Pluym}}, \bibinfo {author} {\bibfnamefont
  {J.~R.}\ \bibnamefont {Wendt}}, \bibinfo {author} {\bibfnamefont {M.~P.}\
  \bibnamefont {Lilly}}, \ and\ \bibinfo {author} {\bibfnamefont {M.~S.}\
  \bibnamefont {Carroll}},\ }\href {\doibase 10.1063/1.4941421} {\bibfield
  {journal} {\bibinfo  {journal} {Appl. Phys. Lett.}\ }\textbf {\bibinfo
  {volume} {108}},\ \bibinfo {pages} {063101} (\bibinfo {year}
  {2016})}\BibitemShut {NoStop}%
\bibitem [{\citenamefont {Studenikin}\ \emph {et~al.}(2012)\citenamefont
  {Studenikin}, \citenamefont {Thorgrimson}, \citenamefont {Aers},
  \citenamefont {Kam}, \citenamefont {Zawadzki}, \citenamefont {Wasilewski},
  \citenamefont {Bogan},\ and\ \citenamefont
  {Sachrajda}}]{Studenikin:2012p233101}%
  \BibitemOpen
  \bibfield  {author} {\bibinfo {author} {\bibfnamefont {S.~A.}\ \bibnamefont
  {Studenikin}}, \bibinfo {author} {\bibfnamefont {J.}~\bibnamefont
  {Thorgrimson}}, \bibinfo {author} {\bibfnamefont {G.~C.}\ \bibnamefont
  {Aers}}, \bibinfo {author} {\bibfnamefont {A.}~\bibnamefont {Kam}}, \bibinfo
  {author} {\bibfnamefont {P.}~\bibnamefont {Zawadzki}}, \bibinfo {author}
  {\bibfnamefont {Z.~R.}\ \bibnamefont {Wasilewski}}, \bibinfo {author}
  {\bibfnamefont {A.}~\bibnamefont {Bogan}}, \ and\ \bibinfo {author}
  {\bibfnamefont {A.~S.}\ \bibnamefont {Sachrajda}},\ }\href {\doibase
  10.1063/1.4749281} {\bibfield  {journal} {\bibinfo  {journal} {Appl. Phys.
  Lett.}\ }\textbf {\bibinfo {volume} {101}},\ \bibinfo {pages} {233101}
  (\bibinfo {year} {2012})}\BibitemShut {NoStop}%
\bibitem [{\citenamefont {Thorgrimsson}\ \emph {et~al.}(2017)\citenamefont
  {Thorgrimsson}, \citenamefont {Kim}, \citenamefont {Yang}, \citenamefont
  {Smith}, \citenamefont {Simmons}, \citenamefont {Ward}, \citenamefont
  {Foote}, \citenamefont {Corrigan}, \citenamefont {Savage}, \citenamefont
  {Lagally}, \citenamefont {Friesen}, \citenamefont {Coppersmith},\ and\
  \citenamefont {Eriksson}}]{Thorgrimsson:2017p32}%
  \BibitemOpen
  \bibfield  {author} {\bibinfo {author} {\bibfnamefont {B.}~\bibnamefont
  {Thorgrimsson}}, \bibinfo {author} {\bibfnamefont {D.}~\bibnamefont {Kim}},
  \bibinfo {author} {\bibfnamefont {Y.-C.}\ \bibnamefont {Yang}}, \bibinfo
  {author} {\bibfnamefont {L.~W.}\ \bibnamefont {Smith}}, \bibinfo {author}
  {\bibfnamefont {C.~B.}\ \bibnamefont {Simmons}}, \bibinfo {author}
  {\bibfnamefont {D.~R.}\ \bibnamefont {Ward}}, \bibinfo {author}
  {\bibfnamefont {R.~H.}\ \bibnamefont {Foote}}, \bibinfo {author}
  {\bibfnamefont {J.}~\bibnamefont {Corrigan}}, \bibinfo {author}
  {\bibfnamefont {D.~E.}\ \bibnamefont {Savage}}, \bibinfo {author}
  {\bibfnamefont {M.~G.}\ \bibnamefont {Lagally}}, \bibinfo {author}
  {\bibfnamefont {M.}~\bibnamefont {Friesen}}, \bibinfo {author} {\bibfnamefont
  {S.~N.}\ \bibnamefont {Coppersmith}}, \ and\ \bibinfo {author} {\bibfnamefont
  {M.~A.}\ \bibnamefont {Eriksson}},\ }\href {\doibase
  10.1038/s41534-017-0034-2} {\bibfield  {journal} {\bibinfo  {journal} {npj
  Quantum Inf.}\ }\textbf {\bibinfo {volume} {3}},\ \bibinfo {pages} {32}
  (\bibinfo {year} {2017})}\BibitemShut {NoStop}%
\bibitem [{Sup()}]{Suppl}%
  \BibitemOpen
  \href@noop {} {\ }\bibinfo {note} {See Supplementary Material for additional
  data, as well as details of theoretical simulation and explanation of
  analysis procedures.}\BibitemShut {Stop}%
\bibitem [{\citenamefont {Ercan}\ \emph {et~al.}()\citenamefont {Ercan},
  \citenamefont {Coppersmith},\ and\ \citenamefont {Friesen}}]{ErcanPrep}%
  \BibitemOpen
  \bibfield  {author} {\bibinfo {author} {\bibfnamefont {H.~E.}\ \bibnamefont
  {Ercan}}, \bibinfo {author} {\bibfnamefont {S.~N.}\ \bibnamefont
  {Coppersmith}}, \ and\ \bibinfo {author} {\bibfnamefont {M.}~\bibnamefont
  {Friesen}},\ }\href@noop {} {\ }\bibinfo {note} {Manuscript in
  preparation}\BibitemShut {NoStop}%
\bibitem [{\citenamefont {Friesen}\ \emph {et~al.}(2006)\citenamefont
  {Friesen}, \citenamefont {Eriksson},\ and\ \citenamefont
  {Coppersmith}}]{Friesen:2006p202106}%
  \BibitemOpen
  \bibfield  {author} {\bibinfo {author} {\bibfnamefont {M.}~\bibnamefont
  {Friesen}}, \bibinfo {author} {\bibfnamefont {M.~A.}\ \bibnamefont
  {Eriksson}}, \ and\ \bibinfo {author} {\bibfnamefont {S.~N.}\ \bibnamefont
  {Coppersmith}},\ }\href {\doibase 10.1063/1.2387975} {\bibfield  {journal}
  {\bibinfo  {journal} {Appl. Phys. Lett.}\ }\textbf {\bibinfo {volume} {89}},\
  \bibinfo {pages} {202106} (\bibinfo {year} {2006})}\BibitemShut {NoStop}%
\bibitem [{\citenamefont {Boykin}\ \emph
  {et~al.}(2004{\natexlab{a}})\citenamefont {Boykin}, \citenamefont {Klimeck},
  \citenamefont {Eriksson}, \citenamefont {Friesen}, \citenamefont
  {Coppersmith}, \citenamefont {von Allmen}, \citenamefont {Oyafuso},\ and\
  \citenamefont {Lee}}]{Boykin:2004p115}%
  \BibitemOpen
  \bibfield  {author} {\bibinfo {author} {\bibfnamefont {T.~B.}\ \bibnamefont
  {Boykin}}, \bibinfo {author} {\bibfnamefont {G.}~\bibnamefont {Klimeck}},
  \bibinfo {author} {\bibfnamefont {M.~A.}\ \bibnamefont {Eriksson}}, \bibinfo
  {author} {\bibfnamefont {M.}~\bibnamefont {Friesen}}, \bibinfo {author}
  {\bibfnamefont {S.~N.}\ \bibnamefont {Coppersmith}}, \bibinfo {author}
  {\bibfnamefont {P.}~\bibnamefont {von Allmen}}, \bibinfo {author}
  {\bibfnamefont {F.}~\bibnamefont {Oyafuso}}, \ and\ \bibinfo {author}
  {\bibfnamefont {S.}~\bibnamefont {Lee}},\ }\href {\doibase 10.1063/1.1637718}
  {\bibfield  {journal} {\bibinfo  {journal} {Appl. Phys. Lett.}\ }\textbf
  {\bibinfo {volume} {84}},\ \bibinfo {pages} {115} (\bibinfo {year}
  {2004}{\natexlab{a}})}\BibitemShut {NoStop}%
\bibitem [{\citenamefont {Boykin}\ \emph
  {et~al.}(2004{\natexlab{b}})\citenamefont {Boykin}, \citenamefont {Klimeck},
  \citenamefont {Friesen}, \citenamefont {Coppersmith}, \citenamefont
  {vonAllmen}, \citenamefont {Oyafuso},\ and\ \citenamefont
  {Lee}}]{Boykin:2004p165325}%
  \BibitemOpen
  \bibfield  {author} {\bibinfo {author} {\bibfnamefont {T.~B.}\ \bibnamefont
  {Boykin}}, \bibinfo {author} {\bibfnamefont {G.}~\bibnamefont {Klimeck}},
  \bibinfo {author} {\bibfnamefont {M.}~\bibnamefont {Friesen}}, \bibinfo
  {author} {\bibfnamefont {S.~N.}\ \bibnamefont {Coppersmith}}, \bibinfo
  {author} {\bibfnamefont {P.}~\bibnamefont {vonAllmen}}, \bibinfo {author}
  {\bibfnamefont {F.}~\bibnamefont {Oyafuso}}, \ and\ \bibinfo {author}
  {\bibfnamefont {S.}~\bibnamefont {Lee}},\ }\href {\doibase
  10.1103/PhysRevB.70.165325} {\bibfield  {journal} {\bibinfo  {journal} {Phys.
  Rev. B}\ }\textbf {\bibinfo {volume} {70}},\ \bibinfo {pages} {165325}
  (\bibinfo {year} {2004}{\natexlab{b}})}\BibitemShut {NoStop}%
\bibitem [{\citenamefont {Faulkner}(1969)}]{Faulkner:1969p713}%
  \BibitemOpen
  \bibfield  {author} {\bibinfo {author} {\bibfnamefont {R.~A.}\ \bibnamefont
  {Faulkner}},\ }\href {\doibase 10.1103/PhysRev.184.713} {\bibfield  {journal}
  {\bibinfo  {journal} {Phys Rev}\ }\textbf {\bibinfo {volume} {184}},\
  \bibinfo {pages} {713} (\bibinfo {year} {1969})}\BibitemShut {NoStop}%
\end{thebibliography}%
\end{document}